%% file: main.tex
\documentclass[sigconf]{acmart}

\AtBeginDocument{%
  }

\include{_macros}

\usepackage{graphicx}
\usepackage{subcaption}
\usepackage{tabularx}
\usepackage{multirow}
\usepackage{pdfpages}
\usepackage{booktabs}
\usepackage{makecell} 
\usepackage{ccicons}

\definecolor{Visualization}{HTML}{F6D6AD}
\definecolor{Questioning}{HTML}{CAABD8}
\definecolor{quote}{HTML}{7B3F00}
\newcommand{\Visualization}{\texttt{\textcolor{black}{\fcolorbox{Visualization}{Visualization}{Viz}}}}
\newcommand{\Questioning}{\texttt{\textcolor{black}{\fcolorbox{Questioning}{Questioning}{Ques}}}}
\newcommand{\iquote}[1]{\textit{\textcolor{quote}{#1}}}
\copyrightyear{2025}
\acmYear{2025}
\setcopyright{cc}
\setcctype{by}
\acmConference[CHI '25]{CHI Conference on Human Factors in Computing
Systems}{April 26-May 1, 2025}{Yokohama, Japan}\acmBooktitle{CHI Conference on Human Factors in Computing Systems (CHI
'25), April 26-May 1, 2025, Yokohama,
Japan}\acmDOI{10.1145/3706598.3714052}
\acmISBN{979-8-4007-1394-1/25/04}
\begin{document}


\title[Are We On Track? AI-Assisted Active and Passive Goal Reflection During Meetings]{Are We On Track? AI-Assisted Active and Passive Goal Reflection During Meetings}



\author{Xinyue Chen}
\authornote{Both authors contributed equally to this research.}
\authornote{The work was done when the co-author was employed at Microsoft.}
\affiliation{%
  \institution{University of Michigan}
  \city{Ann Arbor}
  \country{United States}}
\email{xinyuech@umich.edu}

\author{Lev Tankelevitch}
\authornotemark[1]
\affiliation{%
  \institution{Microsoft Research}
  \city{Cambridge}
  \country{United Kingdom}
}
\email{lev.tankelevitch@microsoft.com}

\author{Rishi Vanukuru}
\authornotemark[2]
\affiliation{%
  \institution{University of Colorado
Boulder}
  \city{Boulder}
  \country{United States}}
\email{rishi.vanukuru@colorado.edu}

\author{Ava Elizabeth Scott}
\authornotemark[2]
\affiliation{%
  \institution{University College London}
  \city{London}
  \country{United Kingdom}
}
\email{ava.scott.20@ucl.ac.uk}

\author{Payod Panda}
\affiliation{%
  \institution{Microsoft Research}
  \city{Cambridge}
  \country{United Kingdom}}
\email{payod.panda@microsoft.com}

\author{Sean Rintel}
\affiliation{%
  \institution{Microsoft Research}
  \city{Cambridge}
  \country{United Kingdom}}
\email{serintel@microsoft.com}

\renewcommand{\shortauthors}{Xinyue Chen, Lev Tankelevitch, Rishi Vanukuru, Payod Panda, Ava Elizabeth Scott, Sean Rintel }

\begin{abstract}
Meetings often suffer from a lack of intentionality, such as unclear goals and straying off-topic. Identifying goals and maintaining their clarity throughout a meeting is challenging, as discussions and uncertainties evolve. Yet meeting technologies predominantly fail to support meeting intentionality. AI-assisted reflection is a promising approach. To explore this, we conducted a technology probe study with 15 knowledge workers, integrating their real meeting data into two AI-assisted reflection probes: a passive and active design. Participants identified goal clarification as a foundational aspect of reflection. Goal clarity enabled people to assess when their meetings were off-track and reprioritize accordingly. Passive AI intervention helped participants maintain focus through non-intrusive feedback, while active AI intervention, though effective at triggering immediate reflection and action, risked disrupting the conversation flow.  We identify three key design dimensions for AI-assisted reflection systems, and provide insights into design trade-offs, emphasizing the need to adapt intervention intensity and timing, balance democratic input with efficiency, and offer user control to foster intentional, goal-oriented behavior during meetings and beyond.
\end{abstract}


\begin{teaserfigure}
  \includegraphics[width=\textwidth]{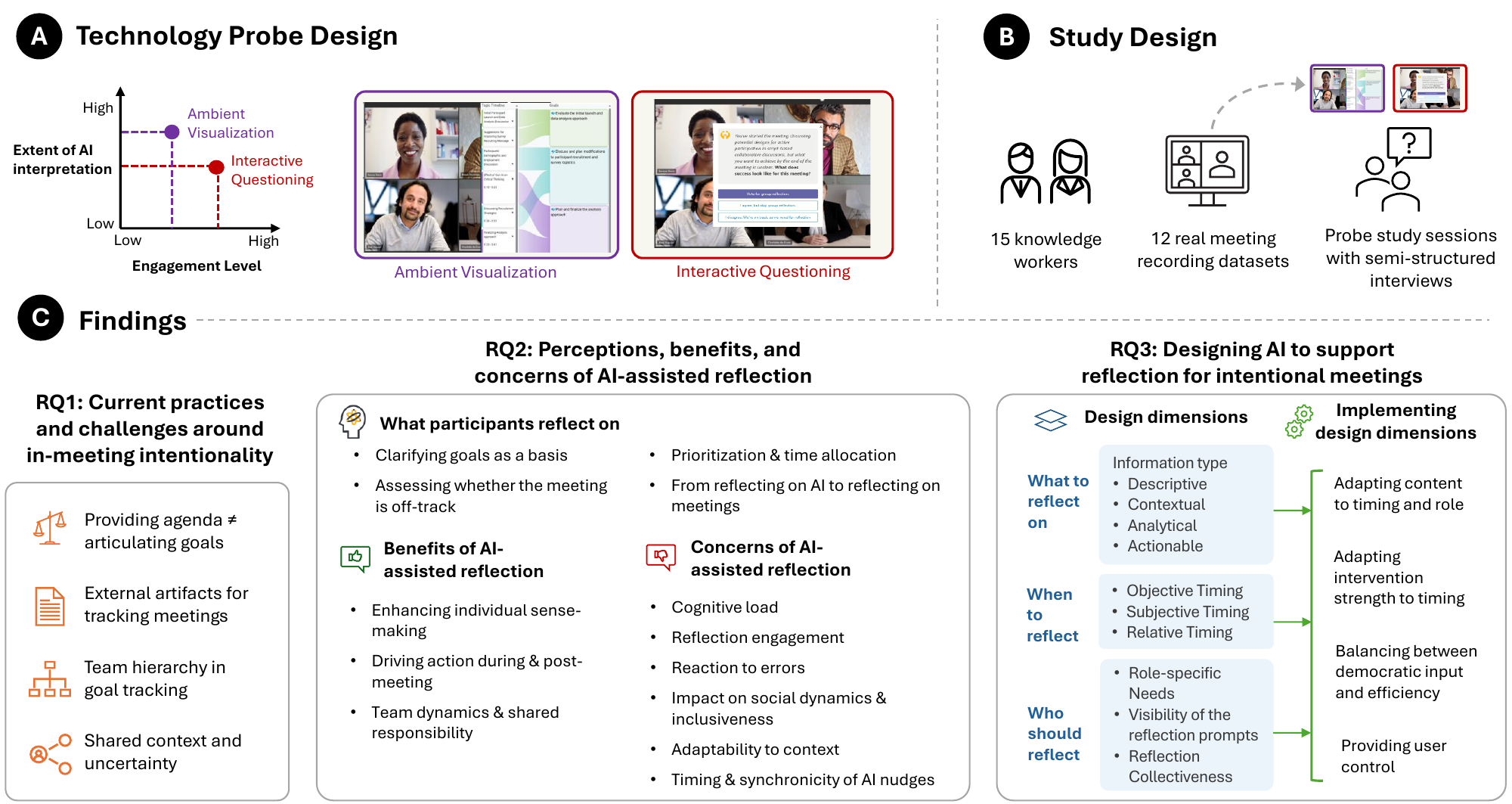}
  \caption{Overview of the study and findings. Based on insights from interviews with participants using two technology probes, we explore how knowledge workers maintain goal alignment in meetings (RQ1), uncover the perceptions, benefits, and concerns of AI-assisted meeting reflection (RQ2), and propose design dimensions and implementation considerations to optimize the role of AI in fostering intentional, goal-oriented behavior during meetings (RQ3).}
    \label{fig:teaser}
\end{teaserfigure}


\maketitle


\section{Introduction}
\label{sec:intro}

While meetings play a crucial role in supporting team planning, coordination, and decision-making \cite{lopez2014working,cao2021large}, they are often criticized for
being ineffective \cite{geimer2015meetings,MicrosoftWorkTrendIndex}. Prior work has explored either micro-behaviors that make a meeting functionally effective (such as problem-solving or action planning \cite{kauffeld2012meetings}) or meeting design characteristics, such as setting agendas, maintaining punctuality, and having dedicated facilitators \cite{leach2009perceived, geimer2015meetings}. Although necessary, they focus on procedure rather than fostering intentionality---knowing why a meeting is occurring and focusing activity on that purpose. This starts with understanding a meeting's goals \cite{scott2024mental}. Assuming that goals can be identified, maintaining goal clarity \textit{throughout} a meeting is challenging, because discussions evolve, and uncertainties or conflicts between personal and collective goals often arise \cite{lopez2022purpose}.

Commercial meeting technologies have historically focused more on features for the transmission of meeting content than intentionality. HCI research has begun exploring the potential of AI-driven tools for real-time feedback during meetings, including visualizing team dynamics \cite{aseniero2020meetcues, chen2023meetscript} and tracking discussion topics \cite{chandrasegaran2019talktraces}. These real-time features improve team awareness but typically only provide surface-level insights without addressing the deeper cognitive processes required to align actions with meeting goals. We argue that meeting technology needs to address the role of \textit{reflection}---examining one's experiences to gain new insights \cite{boud2013reflection}---in promoting workplace intentionality \cite{kocielnik2018designing, wolfbauer2023rebo, fritz2023cultivating}. 

To address this gap, we conducted a design probe study with 15 knowledge workers. We developed two AI-assisted reflection probes: 1) an \textsc{Ambient Visualization} probe, which provides passive, continuous feedback on goal alignment without disrupting the meeting flow, and 2) an \textsc{Interactive Questioning} probe, which actively nudges participants to reflect on whether the current discussion is aligned with the meeting’s objectives at key moments. By integrating real meeting data from participants into these probes, we explored their perceptions of AI-assisted reflection in meetings and examined how to elicit meaningful reflection and encourage intentional behaviors.

Participants identified goal clarification as a foundational value of reflection. They told us that clarifying goals would help them assess when their meetings were off-track and reprioritize accordingly. Passive AI intervention was found to help participants maintain focus through non-intrusive feedback. Active AI intervention, though effective at triggering immediate reflection and action, was found to risk disrupting the meeting's flow. Based on participants'  feedback, we highlight three key dimensions for designing AI-assisted reflection systems: \textit{what} to reflect on (i.e., descriptive, contextual, analytical, and actionable information), \textit{when} to reflect (i.e., objective, subjective, and relative timing), and \textit{who} should reflect (i.e., role-specific needs, visibility, and collectiveness). Participants also provided insights into design trade-offs, emphasizing the need to adapt intervention intensity and timing, balance democratic input with efficiency, and offer user control to foster intentional, goal-oriented behavior during meetings and beyond. As such, our contributions are:

\begin{enumerate}
    \item Empirical findings on how knowledge workers maintain goal alignment in meetings.
    \item Analysis of the benefits and limitations of using AI to support goal reflection during meetings and how these interventions affect potential action during meetings and beyond meetings.
    \item Design dimensions and implementation suggestions that optimize the role of AI in fostering intentional, goal-oriented behavior during meetings, with broader implications for AI-driven meeting support technologies.
\end{enumerate}

\section{Related work}
\label{sec:bg}

\subsection{ Meeting effectiveness and intentionality}
\label{subsec:bg-effectiveness}

While clear goals have been shown to foster shared understanding and improve outcomes \cite{a2014understanding, romano2001meeting}, a significant amount of prior work has focused on more tractable meeting design characteristics, such as agendas \cite{leach2009perceived, perkins2009executive, geimer2015meetings, lopez2022purpose, bang2010effectiveness, cohen2011meeting, a2014understanding}. However, agendas may not always improve meeting effectiveness, as they often emphasize the procedural completion of items on time rather than addressing meetings' underlying priorities \cite{miranda1993impact, garcia2005voting}. 

Recent work has emphasized the importance of `meeting intentionality', involving deliberate actions in planning, monitoring, and adjusting meetings, including setting goals and tracking progress \cite{de2020shared}. Maintaining goal clarity throughout evolving discussions is challenging, requiring monitoring and adjustment of personal and collective objectives \cite{kocsis2015designing} and metacognition---awareness and regulation of one’s cognitive processes \cite{tankelevitch2024metacognitive}. Despite the importance of intentionality, there is limited understanding of how to foster it in meetings.
   
\subsection{Technologies to support intentional meetings }
\label{subsec:bg-intentionaltech}

Early work on Group Support Systems (GSS) aimed to improve decision-making by structuring discussions, idea generation, and voting processes \cite{jerry2000group, conklin2005dialogue, cheng2007framework}. However, GSS often relied on predefined structures, which limited effectiveness when goals were poorly defined \cite{de2003silver}.

Recent work uses AI to provide real-time feedback on team dynamics and improve situational awareness via multi-modal cues \cite{karahalios2009social, kalnikaite2012markup, chen2023meetscript, kelly2016can, aseniero2020meetcues}. Work like TalkTraces visualize discussions in relation to agendas and topics semantically \cite{chandrasegaran2019talktraces, kim2016improving}. Post-meeting tools support indirect meeting intentionality by automating summaries \cite{asthana2023summaries}, tracking action items \cite{sachdeva2021action}, and providing dashboards to monitor effectiveness \cite{shi2018meetingvis, samrose2018automated, samrose2018coco, samrose2021meetingcoach}. Systems like MeetingCoach \cite{samrose2021meetingcoach}, and CoCo \cite{samrose2018automated, samrose2018coco} go further, offering suggestions to help users consider how their behaviors impact group dynamics. Although these tools lack in-meeting functionality, they demonstrate how reflection can be scaffolded to support intentionality. 

Two major limitations persist. First, most tools focus on content, providing insights into discussion patterns but often lacking support to help people understand \textit{why} certain behaviors are necessary and how to adjust their actions in real-time to align with meeting goals. Second, tools have been evaluated primarily in low ecological validity conditions---unfamiliar participants discuss pre-determined tasks in simulated meetings---failing to capture the complexities of real-world workplace meetings. Our study addresses these two limitations.

\subsection{Reflection: A workplace perspective}
\label{subsec:bg-reflection}

The concept of reflection has been applied in various workplace contexts, such as task performance \cite{anseel2009reflection}, time management \cite{whittaker2016don, pammer2015value}, well-being \cite{kocielnik2018designing}, and productivity \cite{rooksby2016personal, yen2017listen}. \citet{schon2017reflective} distinguishes between reflection-in-action, involving real-time adjustments, and reflection-on-action, which occurs post-event. Reflection-in-action has been found to improve performance in work settings \cite{munby1989reflection}. Reflection fosters awareness, which then influences behavior change, similar to the Hawthorne effect \cite{lied1998hawthorne}. Reflection is not solely an individual activity. Team reflection helps align members on roles, discover interaction patterns, and reinforces effective practices \cite{lacerenza2018team, park2023retrospector, fritz2023cultivating}.

Reflection can be promoted \textit{actively}, such as using discussion scripts or scaffolding questions \cite{baker1997promoting, dillenbourg2002over}, or \textit{passively}, such as through subtle data visualizations \cite{sharmin2013reflectionspace, bentvelzen2022revisiting}. Passive systems are often criticized for assuming that reflection naturally occurs upon exposure to information \cite{slovak2017reflective}. Active scaffolding may be more effective \cite{slovak2017reflective}, as emphasized by Fleck and Fitzpatrick’s `levels of reflection' framework \cite{fleck2010reflecting}. However, active approaches interrupt tasks, and workplaces often limit opportunities for such reflection \cite{di2015learning}. \citet{bentvelzen2022revisiting} identified key design resources for supporting reflection, including temporal perspectives such as slowness, which encourages deeper reflection by slowing down user interactions. Reflection is also viewed as a cyclic process, requiring systems that let users navigate and control data interactions at different levels of abstraction \cite{bentvelzen2021technology}. 

Drawing on this work, we consider \textbf{engagement level} as a key design factor in AI-assisted reflection during meetings. We explore AI interventions that are subtle and ambient (passive) or more direct and intermittent (active). Additionally, we consider the level of abstraction when presenting information to promote reflection. 

\subsection{Generative AI in teamwork and team communication}
\label{sec:bg-llms}

Generative AI (GenAI) has led to a surge in tools for teams, focusing on creativity \cite{suh_luminate_2024, liuHowAIProcessing2023}, ideation \cite{xu_jamplate_2024}, decision-making \cite{chiang_enhancing_2024}, and planning \cite{schroder_autoscrum_2023}, and task engagement \cite{arakawa_catalyst_2023}. Commercial meeting systems (e.g., Zoom\footnote{ Zoom’s Meeting AI Companion: \url{https://www.zoom.com/en/ai-assistant/}} and Microsoft Teams\footnote{Microsoft Teams Copilot: \url{https://copilot.cloud.microsoft/en-US/copilot-teams}}) are now leveraging GenAI to enhance meeting efficiency through real-time note-taking, summarization, and action item management. HCI researchers are exploring further ahead to how GenAI can transform meeting workflows. CoExplorer \cite{park2024coexplorer} predicts meeting phases and actively adapts the entire meeting interface for those different phases, while CrossTalk \cite{xia2023crosstalk} suggests context-appropriate actions like screen sharing. Though these tools reduce cognitive load with proactive AI assistance, they risk causing distractions by offering help too frequently or at inappropriate times. This is related to concerns about over-reliance on AI, which can reduce cognitive engagement and critical thinking \cite{liuHowAIProcessing2023}---i.e., the `Assistance Dilemma' phenomenon \cite{koedinger2007exploring}. This dilemma is central to meetings, as too much AI involvement and interpretation can hinder users' \textit{active} engagement and reflection on goals, while too little can cause cognitive overload when users independently parse real-time information  \cite{chen2023meetscript, scott2024mental}. 

Drawing on this research, we explore the \textbf{extent of AI interpretation}: the degree to which AI provides direct feedback data (high AI interpretation) versus guiding users through reflective questions (low AI interpretation), thereby encouraging agency and engagement in reflection \cite{shaer_ai-augmented_2024, wuReflectiveTutoringFramework2011}.

\subsection{Research questions}
\label{subsec:bg-RQs}

Based on the prior work above, we argue that \textit{maintaining intentionality} is key to meeting effectiveness, but the empirical understanding of individuals' practices and how to support intentionality during meetings is limited. Real-time feedback, visualization, and reflection show promise, but most tools are tested in simulated settings, limiting their applicability to real-world meetings. This study explores how AI-assisted reflection can support meeting intentionality through two technology probes and the use of real workplace meeting data. Our research questions were:
\begin{itemize} 
\item RQ1. How do individuals currently maintain intentionality in meetings, and what challenges do they face? 
\item RQ2. What are user perceptions of AI-assisted reflection during meetings? 
\item RQ3. How should AI be designed to support intentional and effective meetings? 
\end{itemize}

\section{Method}
\label{sec:method}

To understand how intentionality is currently supported during meetings, we asked knowledge workers about their meeting practices, analyzed one of their meetings, and then engaged them with two technology probes \cite{hutchinson2003technology, khadpe2024discern} to explore how they would perceive and respond to AI-assisted reflection support during workplace meetings. 
To increase our study's ecological validity, we adapted ideas from the video-stimulated recall method \cite{rowe2009using, morgan2007using}. We asked each participant to provide us with a recording and transcript of one of their real meetings, which we analyzed ourselves but also integrated into the probes to simulate how AI-assisted reflection support would be experienced. While participants interacted with their real meeting in the probes, we used semi-structured interviews to explore their perceptions and considerations for design. 

\subsection{Participants}
\label{subsec:method-participants}

Following ethical approval\footnote{Ethics authorization was provided by Microsoft Research’s Institutional Review Board (IORG0008066, IRB00009672).}, we recruited a purposive sample of 15 employees from various work domains within a global technology company. Employees were recruited through a combination of convenience sampling, snowballing, and batch emails. We aimed for diversity in gender, age, location, work area, seniority, hybrid work status, and managerial roles (see \autoref{tab:demographic}). Participants provided informed consent and were thanked with a gift voucher.

\renewcommand{\arraystretch}{1.25}

\begin{table*}[h]
\centering
\begin{tabular}{llc||llc}
\toprule
\textbf{Dimension} & \textbf{Sub-dimensions}       & \textbf{Participants} & \textbf{Dimension}   & \textbf{Sub-dimensions}   & \textbf{Participants} \\ \hline
\textbf{Age}       & 18-29                         & 4                     & \textbf{Job Level}   & Early Career              & 4                     \\ 
                   & 30-44                         & 7                     &                      & Senior                    & 6                     \\ 
                   & 45-59                         & 4                     &                      & Principal                 & 3                     \\ \cline{1-6}
\textbf{Gender*}   & Man                           & 10                    & \textbf{Work Area}   & Research                  & 11                    \\ 
                   & Woman                         & 5                     &                      & Customer Support          & 2                     \\ 
\cline{1-3}
\textbf{Role}      & Individual Contributor        & 11                    &                      & Product Development       & 1                     \\ 
                   & Manager                       & 4                     &                      & Marketing/Promotion       & 1                     \\ 
\bottomrule
\end{tabular}
\caption{Participant demographics. {*For gender, no participants identified as non-binary or declined to answer.}}
\label{tab:demographic}
\end{table*}

\subsection{Technology probe design}
\label{subsec:method-probedesign}

Probes are widely used in HCI to engage participants early in the design process \cite{wallace2013making}. Technology probes are functional technological artifacts that are “open-ended with respect to use and users
should be encouraged to reinterpret them” \cite{hutchinson2003technology, khadpe2024discern}.
Our two probes aimed to stimulate users' reflective thinking during meetings and gather insights on how AI can foster reflection and support intentionality (RQ2). Additionally, we aimed to surface user considerations on how AI can adapt to the complexities of real-world meetings, including context, timing, and team dynamics (RQ3).

Neither the two dimensions of the probes nor the designs that instantiate them are necessarily ideal scenarios or best possible interfaces. Rather, they are intended to reveal how people might like to be supported in reflective practices during meetings (if at all), and how different meeting contexts might benefit from varying ways of promoting reflection. Similarly, our goal was not to assess the usability of the probes but rather to derive design implications for supporting in-meeting reflection and intentionality.

\subsubsection{Core design dimensions}
\label{subsec:method-coredimensions}

Drawing on research on reflection technology \cite{bentvelzen2022revisiting, boud2013reflection, fleck2010reflecting, schon2017reflective} and the use of large language models (LLMs) in real-time work environments \cite{liuHowAIProcessing2023, schroder_autoscrum_2023, xu_jamplate_2024, shaer_ai-augmented_2024, kazemitabaar_codeaid_2024}, we focused on two dimensions of AI-assisted reflection: \textit{Engagement level} and \textit{Extent of AI interpretation}.

\paragraph{Engagement level:} This dimension focuses on the nature and degree by which the AI system engages users in the reflection process \cite{bentvelzen2021technology, bentvelzen2022revisiting}. As per §\ref{subsec:bg-reflection}, more direct and intermittent nudges have been shown to encourage active reflection, often triggering immediate and focused action from users \cite{park_retrospector_2023, shin_reflection_2017, wuReflectiveTutoringFramework2011}. These nudges can effectively engage users but may disrupt ongoing activities. On the other hand, more passive approaches, such as providing ambient and continuous support, help users maintain an ongoing awareness and track discussions, although they may be overlooked or lead to less active engagement \cite{chandrasegaran2019talktraces, aseniero2020meetcues}. Our two probes were designed to provide contrasting engagement levels: intermittent active calls to attention versus ongoing passive information to which the user might attend. 

\paragraph{Extent of AI interpretation:} This dimension pertains to how much the AI interprets and processes information on behalf of the user.  As per §\ref{sec:bg-llms}, there is a trade-off (the `assistance dilemma'): AI that interprets and directly provides decision-making information can improve awareness of relevant information with minimal user effort \cite{xu_jamplate_2024, liu_selenite_2024}, but can also lead to overreliance, insufficient reflection, and less critical thinking \cite{zhengCompetentRigidIdentifying2023,do_err_2023}. In contrast, AI that merely scaffolds thought processes (with minimal interpretation) can encourage deeper thinking \cite{liuHowAIProcessing2023, xu_jamplate_2024}, but increases users' cognitive load \cite{liuHowAIProcessing2023, xu_jamplate_2024, shaer_ai-augmented_2024}. Our two probes were designed to provide contrasting levels of AI interpretation: high interpretation through presenting ongoing information versus low interpretation through periodic questions.


Considering these dimensions, we designed two technology probes to explore the impact of diametrically opposed levels of AI involvement and user participation on reflective practices during meetings: An \textsc{Ambient Visualization} probe (also noted as \Visualization \ throughout) and an \textsc{Interactive Questioning} probe (also noted as \Questioning \ throughout). 

    
   

\subsubsection{Probe 1: Ambient Visualization}
\begin{figure*}[ht]
    \centering
    \includegraphics[width=1\linewidth]{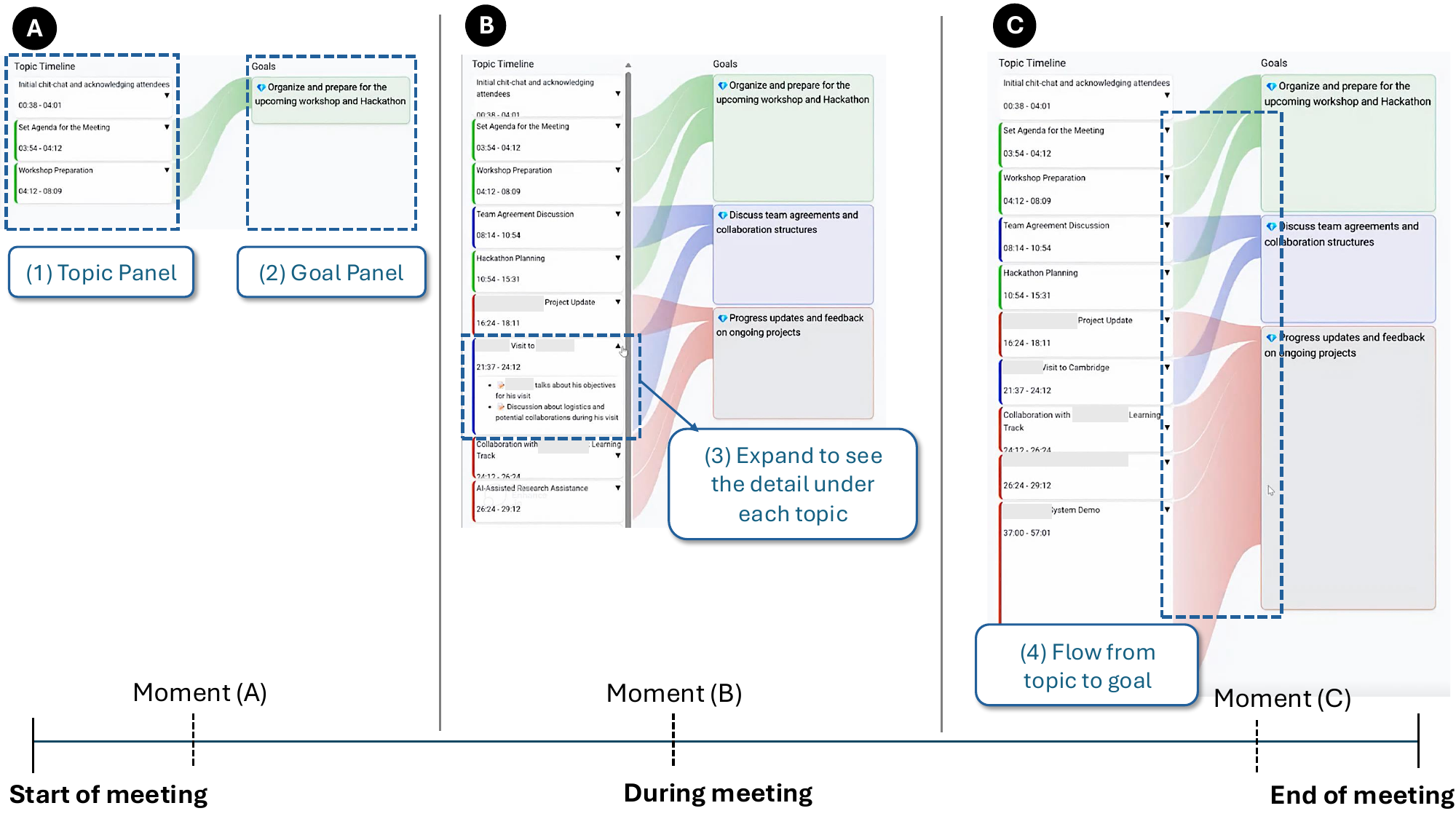}
    \caption{Ambient Visualization: The visualization evolves over time as users observe it at different moments during the meeting (e.g., panels A-C), with the content becoming richer as the discussion progresses}
    \label{fig:Ambient}
\end{figure*}

The \textsc{Ambient Visualization} probe (\Visualization) (\autoref{fig:Ambient}) exemplifies a combination of \textit{high} extent of AI interpretation and \textit{low} engagement level. It supports users by continuously analyzing and displaying meeting topics and goals, providing a high level of interpretative assistance. However, as this support is continuous and ambient, users are not required to react immediately, and thus the assumed level of engagement is low.

We designed this passive probe to infer and display meeting goals, conversation topics, and their relationship (i.e., how topics contribute to goals), using real-time GenAI-driven transcript analysis. Integrated into the right sidebar of a meeting interface, topics are arranged chronologically on the left, while goals are displayed on the right as they are identified, each assigned a random color. To help users evaluate whether discussions align with goals, a color-coded flow connects related topics to goals, helping users quickly identify relevant information \cite{chen2023meetscript, shi2018meetingvis}.
Users can click on a topic to access more detailed information as needed. This aligns with research advocating for multiple levels of abstraction to enhance reflection \cite{bentvelzen2022revisiting}. As illustrated in \autoref{fig:Ambient}, the visualization evolves in real-time, providing a dynamic, ongoing representation of AI-inferred goals and topics, presented with minimal interruption or interactivity. 

\subsubsection{Probe 2: Interactive Questioning}
\begin{figure*}[ht]
    \centering
    \includegraphics[width=\linewidth]{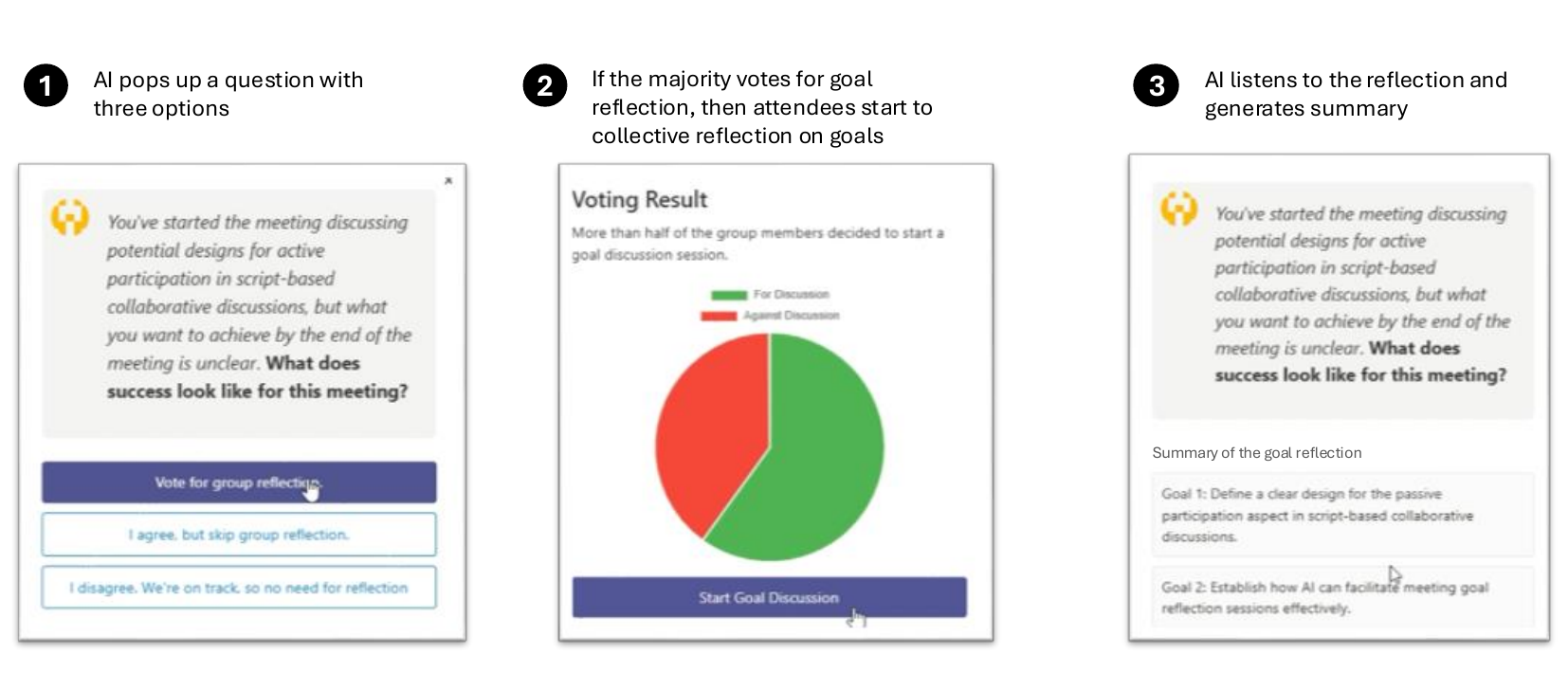}
    \caption{Interactive Questioning: (1) AI pops up a question with three options at the time when AI identifies a reflective discussion is needed. (2) If the majority votes for reflection, then it proceeds. (3) AI listens to and summarizes the goal-oriented reflective discussion.}
    \label{fig:Interactive}
\end{figure*}

The \textsc{Interactive Questioning} probe (\Questioning) (\autoref{fig:Interactive}) exemplifies a combination of \textit{low} extent of AI Interpretation and \textit{high} engagement level. It nudges users with questions at key moments during the meeting, encouraging active reflection. Since the AI poses questions rather than presenting information, it offers an assumed low degree of interpretation. 

We designed this active probe to intervene in two specific conditions: (1) when no clear meeting goals have been established within the first five minutes of the meeting, and (2) when the AI detects that the discussion is drifting away from the main goal of the meeting that AI inferred from the speech. When either condition is met, the AI presents participants with a reflection question tailored to the meeting. For condition (1), it was approximately: \textit{``You've started the meeting discussing [topic X] but what you want to achieve by the end of the meeting is unclear. \textbf{What does success look like for this meeting?}''}. For condition (2), it was approximately: \textit{``You've discussed [topic X] for a while now but the overarching goal of the meeting is [goal Y]. \textbf{Does the current discussion align well with the meeting goal? What can we do to ensure the meeting is on track?}''}. The nudge also presented three response options: (i) \textit{`Vote for group reflection'}, (ii) \textit{`I agree, but skip group reflection'}, or (iii) \textit{`I disagree. We're on track, so no need for reflection'}. The questions are not meant to cover all situations. Instead, they represent important instances where intentionality may be unclear or misaligned. This approach allows participants to experience how they might respond to AI-generated nudges for reflection at specific moments during meetings. Participants then vote on whether or not the team should reflect on the question or continue the meeting. The voting mechanism is included to explore how \textit{group reflection} could be initiated. If the majority of participants vote in favor of group reflection, the AI listens to the ensuing discussion and provides a summary. This probe deliberately introduces AI in an intrusive manner, with the goal of actively promoting reflection within the meeting. 

\subsubsection{System implementation}

We developed working prototypes of the two probes. Their backend is built with Node.js, while the frontend features a simulated meeting interface that displays meeting recordings. During video playback, each turn's transcript is sent to the backend for real-time processing.

In the passive probe (\Visualization), GPT-4 detects topic changes in real-time by analyzing each turn against previous ones. When a change is identified, it summarizes key points, updates the topic panel, and identifies emerging goals for the goal panel. When new topics or goals arise, the AI assesses the relationships between them and updates the visual flow using D3.js. Instead of a post-meeting summary, this workflow simulates AI's real-time identification of topics, goals, and relationships during the meeting.

In the active probe (\Questioning), one AI agent analyzes the transcript to evaluate whether the discussion aligns with goals or has drifted off track. If goals are absent or the discussion deviates, the AI dynamically determines the appropriate timing to introduce a reflection question. If so, another agent generates the question in real-time. Both the timing and content of the questions are context-based and dynamic.  Participants can vote on the question and view a simulated voting result and reflective discussion summary. Detailed system prompts are provided in the supplementary materials.

\subsection{User feedback via video-stimulated recall}

To increase the ecological validity of our probe study (RQ2 and RQ3), we integrated real meeting data (recording and transcript) from each participant into the probes. We encouraged participants to imagine they were actually in the meeting rather than watching a recording and to imagine how they would react to these AI-assisted nudges in a live setting. This method is an adaptation of video-stimulated recall (VSR) common in educational research \cite{rowe2009using, morgan2007using,nguyen2013video}, albeit instead of asking participants to reflect on their past behavior, they are asked to consider how they would respond in a counterfactual scenario involving our probes. VSR ``helps participants retrospectively articulate their thought processes by minimizing self-consciousness, by maximizing their psychological immersion in the activity preceding the interview, and by triggering memories of these cognitive processes'' \cite{audetat2021understanding}. As participants interacted with the probes, we used a semi-structured interview protocol to ask about their thoughts.

Our approach strikes a balance between lab studies of simulated meetings \cite{spittle2024comparing, park2024coexplorer}, which are far removed from the rich and dynamic discussions of real workplace meetings, and the real-time deployment of probes during real meetings \cite{aseniero2020meetcues}, which, although valuable, has practical and privacy challenges. Furthermore, given our focus on reflection, introducing probes into live meetings would not effectively capture participants' immediate responses without interfering with the meeting. Thus, our approach also enabled capturing participants' rich responses as they arose. 

We acknowledge that, despite being grounded in real data, our approach does not fully replicate the dynamics of live meetings. Nevertheless, we observed that our probes provided sufficient context to stimulate meaningful reflection from participants on how AI might support intentionality during meetings, for example, participants often verbalized their thoughts with phrases like, \textit{``if I were in this meeting and I saw this, I would...''}. 

Additionally, collecting real meeting data also helped us better understand participants' current practices during meetings (RQ1). It provided behavioral insights into how intentionality is maintained in real-world settings and enabled us to ask more personalized questions during interviews.  

\subsection{Study procedure}
\begin{figure*}[htbp]
    \centering
    \includegraphics[width=1\linewidth]{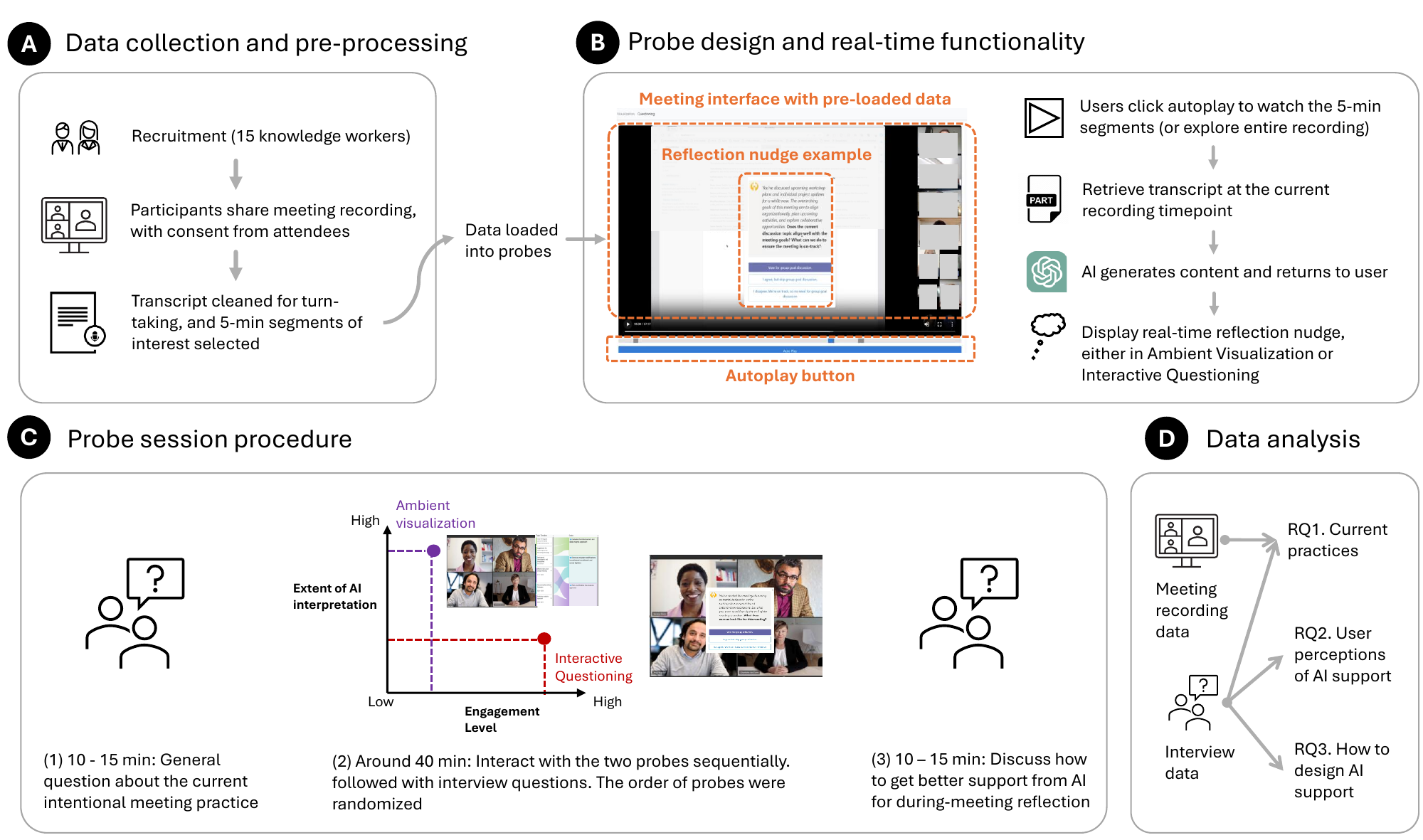}
    \caption{Study overview. The diagram illustrates the overall study workflow, including data-preprocessing (A),  system setups, and the workflow of simulating real-time AI-assisted reflection during the session (B), probe session procedure (C), and data analysis (D).}
        \label{fig:study-overview}
\end{figure*}

Before the session, participants consented and completed an onboarding survey that collected demographic information. Participants also donated meeting data with consent from all attendees (\autoref{fig:study-overview}A). The donated meeting data was pre-processed and inserted into the two AI-assisted prototypes that participants would interact with during the session (\autoref{fig:study-overview}A-B). To focus the sessions on key moments of interest in each meeting, the researcher chose 2-3 segments (approximately 5 minutes each) in each meeting. Segments were chosen based on pre-session testing of the AI, focusing on moments where the AI identified new topics and goals (Probe 1) or determined that a reflection question was appropriate to present (Probe 2). 

Study sessions were conducted remotely through a videoconferencing platform. Participants were asked to share their screens during the session. Each session lasted approximately 60 minutes and was recorded for analysis.

During the session, participants first answered general questions about how they communicate and track goals and agendas during meetings, and the challenges in staying aligned with those goals (\autoref{fig:study-overview}C). Participants were then introduced to one of the two probes with a demo video (with order counterbalanced across participants). They were guided to interact with each probe, which auto-played the pre-selected segments from their actual meetings to simulate real-time AI intervention. Participants were also able to freely explore the entire meeting using the seek bar in the video player (\autoref{fig:study-overview}B). As participants interacted with each probe, they were asked to consider how they might respond to the AI-assisted reflection support in their own meetings. 

After interacting with both probes, participants answered more general questions about how AI might be integrated into meetings to promote reflection, focusing on aspects such as the timing of interventions and how the AI could support goal alignment and intentionality.

\subsection{Data analysis}
\renewcommand{\arraystretch}{1.0}

\renewcommand{\arraystretch}{1.1} 
\begin{table*}[h]
\centering
\begin{tabular}{p{0.5cm}p{0.5cm}p{5.5cm}p{1cm}p{7cm}}
\toprule
\textbf{ID} & \textbf{Size} & \textbf{Meeting Type} & \textbf{Length} &\textbf{ Participant and role } \\ \hline
M1 & 2 & Recurring; Daily Project Meeting & 37 min  & P1 (active participant)\\ \hline
M2 & 4 & One-time; Project Meeting & 55 min  & P2 (active participant)  \\ \hline
M3 & 3 & Recurring; Weekly Project Meeting & 1h & P3 (passive participant)\\ \hline
M4 & 4 & Recurring; Weekly Team Sync & 30 min  & P4 (passive participant)\\ \hline
M5 & 4 & One-time; Managerial Mentoring Session & 1h  & P5 (organizer), P6 (passive participant) \\ \hline
M6 & 20 & Recurring; Weekly Team Sync & 1h  & P7 (passive participant), P8 (active participant)  \\ \hline
M7 & 9 & Recurring; Weekly Team Sync & 44 min  & P9 (passive participant) \\ \hline
M8 & 3 & One-time; Project Meeting & 1h & P10 (passive participant) \\ \hline
M9 & 5 & Recurring; Weekly Project Meeting & 1h & P11 (organizer) \\ \hline
M10 & 8 & One-time; Onboarding Meeting & 1h  & P12 (organizer)   \\ \hline
M11 & 4 & Recurring; Weekly Team Sync & 33 min & P13, P15 (passive participant)\\ \hline
M12 & 24 & Recurring; Weekly Team Sync & 45 min & P14 (active participant) \\ \bottomrule
\end{tabular}
\caption{Meeting Data Overview. Meeting types are classified based on information provided by users in the consent form. Participants’ roles are classified into three types: organizer, active participant, and passive participant, as identified in the findings of §\ref{subsubsec:findings-design-who} }
\label{tab:recording}
\end{table*}

The analysis focused on both the interview and real meeting data (\autoref{fig:study-overview}D).\footnote{We have fewer meetings (n = 12) than participants (n = 15) because several participants attended the same meetings.} We conducted qualitative data analysis using thematic analysis \cite{braun2012thematic}. First, we applied open coding to identify key themes related to participants' current meeting practices and their reactions to AI-assisted reflection support in meetings. Initial coding was performed by the primary author, and emerging codes were refined in collaboration with the research team. Second, we integrated an additional layer of analysis by coding the meeting data provided by participants (see \autoref{tab:recording} for an overview of the meeting data). This allowed us to annotate specific behaviors and discourse patterns related to goal communication, monitoring, and adjustments during meetings. These coded behaviors were then mapped against the interview data to provide a richer answer to RQ1. The codebook of the meeting data is shown in Appendix \ref{sec: codes}. 

\section{Findings (RQ1): Current practices and challenges of meeting intentionality during meetings }
\label{sec:findings-currentpractices}

We present our findings addressing RQ1 based on our analysis of participants' meeting data and their responses during the study session. 
Each theme presents current practices that participants commonly follow to support intentionality and the challenges they encounter while navigating these practices (\autoref{tab:themes_subthemes}).

\renewcommand{\arraystretch}{1.25} 
\begin{table*}[h]
\centering
\begin{tabular}{p{0.4\textwidth} p{0.55\textwidth} }
\toprule
\textbf{Theme} & \textbf{Sub-theme} \\ \hline
\multirow{3}{*}{\parbox{0.4\textwidth}{\textit{Providing agendas \(\neq\) articulating goals}}} 
& Distinguishing between agendas and goals (Practice) \\ \cline{2-2}
& Listing agendas without articulating goals (Practice) \\ \cline{2-2}
& Agendas cannot ensure intentionality (Challenge) \\ \hline
\multirow{2}{*}{\parbox{0.4\textwidth}{\textit{External artifacts for tracking meetings}} }
& External artifacts for structuring and tracking discussions (Practice) \\ \cline{2-2}
& Difficulty revisiting external references during high-paced discussions (Challenge)  \\ \hline
\multirow{2}{*}{\parbox{0.4\textwidth}{\textit{Team hierarchy in goal tracking}}} 
& Proactive role of managers/organizers (Practice) \\ \cline{2-2}
& Constrained contribution from non-leadership roles   (Challenge) \\ \hline
\multirow{2}{*}{\parbox{0.4\textwidth}{\textit{Shared context and uncertainty in goal-setting}}} 
& Tacit shared knowledge reduces need for explicit goals (Practice) \\ \cline{2-2}
& Evolving nature and uncertainty of meeting goals (Challenge) \\ 
 \bottomrule
\end{tabular}
\caption{Themes and sub-themes of RQ1 findings: \textit{Current practices and challenges of meeting intentionality during meetings}}
\label{tab:themes_subthemes}
\end{table*}

\subsection{Providing agendas \(\neq\) articulating goals}
\label{subsec:findings-agendas}

Meeting goals and agendas are distinct: whereas goals set the destination for a meeting, agendas provide a path to get there \cite{scott2024mental}. Participants themselves often distinguished between agenda and goals in interviews \pid{(P2, P4, P7, P8, P11, P12, P14, P15)}, seeing “Agendas as the spectrum of things that should be discussed" and "goals as outcomes that should be achieved.” \pid{(P7)}. Although many teams had agendas---either through external representations \pid{(5 of 12 meetings)} or verbal descriptions \pid{(8 of 12 meetings)}---fewer articulated specific goals for their meetings \pid{(4 of 12 meetings)}, consistent with prior work \cite{scott2024mental}. Goals were often left unstated due to their dynamic nature and associated uncertainties (see also §\ref{subsec:findings-uncertainty}). 

\begin{smallquote}
    \iquote{“A meeting might be about discussing an issue, but you may not even know if it’s possible to come to a conclusion.”} \pid{(P7)}
\end{smallquote}

Having an agenda does not necessarily provide clarity on time management, nor does it ensure that the meeting will be effective. Of the five meetings with explicit agendas, none allocated time for each item, and time management behaviors typically emerged only toward the end, when participants realized they were running out of time. With only agendas but without clear goals, participants struggled to prioritize discussions \pid{(P1, P3)}, often leading to overly long discussions \pid{(6 of 12 meetings)} or under-addressed topics \pid{(4 of 12 meetings)}.

\begin{smallquote}
    \iquote{“We just list what needs to be discussed. When you talk about data sets or specific experiments, you can easily lose track of time.”} \pid{(P1)}
\end{smallquote}

\subsection{External artifacts for tracking meetings}
\label{subsec:artifacts}

External artifacts, such as shared documents and slides, were used to help with meeting tracking \pid{(7 out of 12 meetings)}. 
This was particularly common in larger meetings \pid{(M6, M10)}, where shared documents listed the agenda and discussion points, and participants were asked to add further items. Despite the frequent use of these tools, participants still encountered challenges in tracking discussions, e.g., in referring back to artifacts during high-paced discussions. This was echoed in the meeting data analysis, where 3 out of 12 meetings were found to be off-target. This highlights a gap where external artifacts alone may not sufficiently support effective meeting tracking.

\begin{smallquote}
    \iquote{“In this meeting, my manager usually puts a top slide with the agenda there. But I haven't been as good about circling back to say, `OK. We've addressed this and this during the meeting.'”} \pid{(P8)} 
\end{smallquote}

\subsection{Team hierarchy in goal tracking}
\label{subsec:findings-hierarchy}

 In meetings with clearly listed goals \pid{(4 of 12 meetings)}, managers and organizers took a more proactive approach, intervening more frequently to keep the meeting on track, managing time, prioritizing tasks, gathering input, and ensuring that objectives were aligned with all participants. In contrast, junior participants and non-organizers often felt less responsible for keeping the meeting on track \pid{(P1, P2, P3, P4, P7, P9)}. Despite recognizing when discussions diverged from the agenda, they hesitated to speak up due to social dynamics or fear of overstepping their role. 

\begin{smallquote}
    \iquote{“I noticed that we were diverging from the topic, but since I wasn’t leading the meeting, I didn’t feel it was my place to bring it up.” } \pid{(P5)}
\end{smallquote}

\subsection{Shared context and uncertainty in goal-setting}
\label{subsec:findings-uncertainty}

There were nuances in how participants perceived goals depending on the type of meeting. First, in certain types of meetings where tacit shared knowledge exists, participants often felt that explicit goal-setting was unnecessary. This was especially true for recurring meetings where participants already understood the general objectives without needing formal goal articulation  \pid{(P4, P7, P8, P9, P13, P15)}. 

Second, the evolving nature of meeting goals was emphasized by participants. Echoed in the meeting data, we found that in 3 out of 12 meetings, goals only became defined during the conversation, while 7 meetings included emerging topics that shifted the direction of discussions. Participants highlighted the importance of recognizing such inherent uncertainty and the evolving nature of meeting goals and were skeptical that this uncertainty could be uncovered with predefined goals \pid{(P5, P6, P12)}.

\begin{smallquote}
    \iquote{“The goal is usually very general. We were just trying to understand each other. But I should say it would be helpful if there is a tangible outcome in a way.”} \pid{(P9)}
\end{smallquote}

\section{Findings (RQ2): Perceptions, benefits, and concerns of AI-assisted reflection during meetings}
\label{sec:findings-perceptions}

Our findings from RQ1 highlight the inherent challenges in maintaining meeting intentionality. These obstacles validate the need for interventions to support meeting intentionality.  In the following sections, we explore what participants reflect on (§\ref{subsec:findings-what}) when using reflection probes to support meeting intentionality, the benefits of these methods (§\ref{subsec:findings-benefits}), and concerns about integrating these reflection technologies into their meetings (§\ref{subsec:findings-concerns}).

\subsection{What participants reflect on}
\label{subsec:findings-what}

Participants interacted with the probes and shared how AI assistance might influence their thinking and actions during real meetings. Their reflections focused on clarifying goals, assessing alignment with those goals, prioritizing topics and time management, and reflecting on AI content. We compare responses between the passive probe (\Visualization) and the active probe (\Questioning).

\begin{table*}[ht]
\centering
\begin{tabular}{ll}
\toprule
\textbf{Theme} & \textbf{Sub-theme} \\ \hline
\multirow{2}{*}{\textit{Clarifying goals as a basis}} 
& Recalibrating and structuring goals (\Questioning) \\ \cline{2-2}
& Interpreting and building on AI goals (\Visualization) \\ \hline
\multirow{2}{*}{\textit{Assessing alignment with meeting goals}} 
& Questioning relevance of their own contributions (\Questioning)\\ \cline{2-2}
& Recognizing off-track discussions (\Visualization) \\ \hline
\multirow{2}{*}{\textit{Prioritization and time allocation}} 
& Prioritizing actions based on goal importance (\Questioning) \\ \cline{2-2}
& Reflecting on general discussion structure and time use (\Visualization) \\ \hline
\multirow{2}{*}{\textit{Reflecting on AI content}} 
& Evaluating AI's accuracy and reasoning (\Questioning) \\ \cline{2-2}
& Reflecting on communication clarity due to AI misalignment (\Questioning) \\ \bottomrule
\end{tabular}
\caption{Themes and sub-themes on what participants reflect on}
\label{tab:themes_subthemes_reflections}
\end{table*}

\subsubsection{Clarifying goals as a key value}
\label{subsubsec:findings-clarifying}

Clarifying goals emerged as the primary reaction and value for most participants for both probes \pid{(11/15)}. 
With active reflection (\Questioning), participants frequently began by asking, \textit{"What are we trying to achieve?"} \pid{(13/15)}. 
Goal clarification was not merely superficial; some participants dissected and categorized the various layers of meeting objectives and articulated their logical connections \pid{(P1, P4, P12)}. 

\begin{smallquote}
    \iquote{"So the goal for this stand-up, in the logistic sense, is definitely to update what everyone is working on. We also have a specific research-wise goal ... there is a list of pending questions..."} \pid{(P4, \Questioning)} 
\end{smallquote}

Under passive reflection (\Visualization), the majority of participants \pid{(10/15)} initially responded by interpreting the AI-generated outputs rather than coming up with goals themselves.

\begin{smallquote}
\iquote{"Ah, 'handling reward'... 'feedback and discrepancies'... That's cool... these are the things we want to talk about"} \pid{(P5, \Visualization)}.
\end{smallquote}

Beyond simple interpretation, the ambient visualization nudged participants to build upon the AI's suggestions. Triggered by the AI-identified goals, some participants noticed previously implicit and unplanned goals that emerged during the meeting \pid{(P3, P9, P12, P13)} and occasionally identified goals that the AI failed to define \pid{(P2, P3, P6, P8)}.  




\subsubsection{Assessing alignment with meeting goals}
\label{subsubsec:findings-alignment}

With both probes, all participants evaluated whether their actual discussion aligned with the meeting’s goals. 
Under passive reflection (\Visualization), participants were more likely to recognize and reflect on off-track discussions that they had not previously noticed \pid{(P3, P8, P9, P12)}. 

\begin{smallquote}
\iquote{" I didn't feel like anything was off-track before. But when I look at the visualization, this topic seems a little independent... 
} \pid{(P9, \Visualization)}
\end{smallquote}

In contrast, active reflection (\Questioning) nudges led participants to question the alignment of their personal contributions with the broader team goals \pid{(P2, P4, P5, P9)}. 

\begin{smallquote}
    \iquote{"Okay, probably this issue is only something I and another person have; we can just discuss it offline since it is not related to everyone's goal for this meeting." } \pid{(P5, \Questioning)} 
\end{smallquote}

\subsubsection{Prioritization and time allocation}
\label{subsubsec:findings-prioritization}

Another common theme that people reflected on was the hierarchy of different topics and goals \pid{(11/15)}. 
In the active probe (\Questioning), some participants claimed that they would prioritize their actions based on the perceived importance of different goals \pid{(P3, P5, P6, P8, P9, P11, P15)}. This led to real-time behavioral adjustments and shifts in meeting focus. 

\begin{smallquote}
\iquote{"If I saw this in my meeting, I would consult with my team immediately to determine if she should speak first, given the closer relevance of their content, which might facilitate our decision, rather than me commencing as originally intended.  
} \pid{(P9, \Questioning)}
\end{smallquote}

In contrast, when nudged by the passive visualization (\Visualization), participants were more likely to reflect on the overall structure of the agenda and the time distribution \pid{(P3, P4, P8, P13, P14)}. This reflection did not always lead to immediate adjustments but offered a broader awareness of how much attention was given to various topics.

\begin{smallquote}
    \iquote{ "I think one thing that I didn't notice  was that we were spending a lot of time on determining the computation...
which I don't think was actually super important here."} \pid{(P3, \Visualization)} 
\end{smallquote}

\subsubsection{Reflecting on AI content}
\label{subsubsec:findings-content}

People also reflected on the AI content or nudge itself, which triggered further reflection on the meeting \pid{(P1, P2, P5, P6, P14)}. This pattern was particularly evident in the active probe (\Questioning), where AI-generated questions or nudges highlighted discrepancies or nuances that participants felt compelled to address. For instance, when the AI's defined goals differed from the users' understanding, P5 questioned it: 

\begin{smallquote}
\iquote{``Why did I go there? Why did the AI pick that up?''} \pid{(P5, \Questioning)}
\end{smallquote}

Trying to understand the AI-generated question nudged participants to reflect on their own performance. 

\begin{smallquote}
\iquote{``Maybe I'll think more critically, maybe I may or may not actually convey that clearly in the form,  so that AI didn't really catch the nuances I want to convey''.} \pid{(P2, \Questioning)}
\end{smallquote}

\subsection{Benefits of AI-assisted reflection for meetings}
\label{subsec:findings-benefits}

Across both probes, participants acknowledged the overall value of AI in supporting these aspects, though nuances in how the benefits were realized differed slightly between the two approaches, as illustrated in \autoref{tab:benefits_ai_reflection}. Those benefits centered around \textit{individual sense-making}, enabling \textit{action}, shifting \textit{team dynamics and responsibility}.

\begin{table*}[h!]
\centering
\begin{tabular}{ll}
\toprule
\textbf{Theme} & \textbf{Sub-theme} \\ \hline
\multirow{2}{*}{Enhancing individual sense-making} 
& Surfacing and clarifying meeting objectives \\ \cline{2-2}
& Encouraging ownership of participation \\ \hline
\multirow{4}{*}{Driving action during and post-meeting} 
& Non-intrusive nudging of steady behavior change  (\Visualization) \\ \cline{2-2}
& Nudging of post-meeting review and follow-up  (\Visualization) \\ \cline{2-2}
& Encouraging proactive behavior adjustments in real-time  (\Questioning) \\ \cline{2-2}
& Influencing behavior change with long-term practice (\Questioning) \\ \hline
\multirow{2}{*}{Shaping team dynamics and responsibility} 
& Fostering shared responsibility \\ \cline{2-2}
& Acting as a neutral mediator to reduce affront\\ \bottomrule
\end{tabular}
\caption{Themes and sub-themes of benefits of AI-assisted reflection.}
\label{tab:benefits_ai_reflection}
\end{table*}

\subsubsection{Enhancing individual sense-making} 
\label{subsubsec:findings-sensemaking}

AI-assisted reflection, whether active or passive, played a crucial role in making implicit thoughts and goals explicit, thus enhancing individual sense-making of the meeting. Both probes helped participants surface, clearly define, and understand the meeting's objectives \pid{(9/15)}, and encouraged them to take ownership of their involvement \pid{(12/15)}. 

\begin{smallquote}
 \iquote{ "Normally you have like some goals in your mind, and then 
 the AI summarized the goals from your discussion, then you can sort of like calibrate your thoughts on what actually the goals are... 
 "\pid{(P4, \Visualization)}}.
 \end{smallquote}

\subsubsection{Driving action during and post-meeting}
\label{subsubsec:findings-drivingaction}

AI-assisted reflection also drove tangible (intended) actions both during and after meetings. Passive reflection (\Visualization) was thought of as a non-intrusive layer of information, helping participants become aware of time imbalances or overlooked topics,  providing a steady influence on the meeting's flow \pid{(P1, P7, P9, P10, P11, P12)}.

\begin{smallquote}
\iquote{``Always displaying information for people to make decisions on actually can trigger changes of behaviors at many more points throughout the meeting by keeping in mind that `are we on track? Do I want to add anything for us to discuss?'''} \pid{(P14, \Visualization)}
\end{smallquote}

Additionally, passive visualization was found to be valuable for post-meeting review and follow-up, with the AI-generated artifacts serving as reference points after the meeting \pid{(P5, P6 P12, P13)}.

Participants who were actively nudged (\Questioning) to reflect on their goals were more likely to express an intention to take action in real-time, including reminding others to steer the conversation back on track \pid{(P5, P8, P9, P15)}, prioritizing the agenda items that need discussing in the remaining time \pid{(P3, P4)}, and ensuring all team members have opportunities to speak before the meeting ends \pid{(P2, P6, P13)}. The perceived impact of active reflection also extended beyond the current meeting. Participants saw the potential to adjust their behavior proactively as they became more accustomed to communicating and tracking goals through AI cues \pid{(P3, P5, P7, P11)}.  

\begin{smallquote}
\iquote{``
When you start a new meeting where people are not really sure what it is all about, they might be able to be guided to start with a list of goals over time in this series while being asked to discuss the goals at first several times.''} \pid{(P7, \Questioning)}
\end{smallquote}

\subsubsection{Team dynamics and shared responsibility}
\label{subsubsec:findings-team-dynamics}

AI-assisted reflection was also thought to help shape team dynamics and foster a sense of shared responsibility during meetings \pid{(9/15)}. 

\begin{smallquote}
\iquote{" 
Instead of one person bearing the responsibility to interrupt, 
 the system fosters a shared awareness among everyone knowing there are still three topics to cover."} \pid{(P8, \Visualization)}
\end{smallquote}

One reason for this is that AI seemed to serve as a mediator in discussions \pid{(P2, P4, P6, P11, P13)}. Unlike human interventions, which may carry personal biases or lead to conflicts, AI intervention encouraged participants to reconsider points without the risk of personal affront. 

\begin{smallquote}
\iquote{"The AI prompted us to reconsider a point without anyone feeling attacked." } \pid{(P6, \Questioning)}
\end{smallquote}

Despite its advantages for fostering team dynamics and shared responsibility, there are also concerns about how AI-assisted reflection might affect team interactions, as will be reported in \S\ref{subsubsec:findings-social}.

\subsection{Concerns about AI-assisted reflection during meetings}
\label{subsec:findings-concerns}

Participants acknowledged the benefits of the two AI-assisted probes in fostering reflection and improving meeting dynamics (\S\ref{subsec:findings-benefits}) but also highlighted concerns, as shown in \autoref{tab:concerns}. These concerns included both probe-specific challenges and shared issues, informing design considerations for effective reflective technologies in meetings (\S\ref{sec:findings-design}).

\begin{table*}[h!]
\centering
\begin{tabular}{ll}
\toprule
  \textbf{Theme} & \textbf{Sub-theme} \\ \hline
\multirow{2}{*}
 {Cognitive load} & Overload from continuous visual data (\Visualization) \\ \cline{2-2}
  & Insufficient information from nudges (\Questioning) \\ \hline
\multirow{2}{*}
 {Reflection engagement} & Limited engagement with passive visualization (\Visualization) \\ \cline{2-2}
  & Disruption of discussion flow from active questioning (\Questioning) \\ \hline
  \multirow{2}{*}
 {Reaction to errors} & Over-reliance and overlooked nuances (\Visualization) \\ \cline{2-2}
  & Sensitivity to inaccuracies and frustration (\Questioning) \\ \hline
\multirow{2}{*}{
 Social dynamics} & Hesitancy among individual contributors to share ideas \\ \cline{2-2}
 & Concerns about inclusiveness and manager decision-making \\ \hline
 Adaptability to Context & Challenges in handling diverse meeting types \\ \hline
\multirow{2}{*}{Timing and synchronicity} & Delayed nudges reducing real-time relevance (\Questioning) \\ \cline{2-2}
 & Outdated visual displays limiting timely reflection (\Visualization) \\ 
\bottomrule
\end{tabular}
\caption{Themes and sub-themes of concerns about AI-assisted reflection.}
\label{tab:concerns}
\end{table*}

\subsubsection{Cognitive load: Overload vs. insufficient information}
\label{subsubsec:findings-overload}

Some participants \pid{(5/15)} noted that the continuous stream of visual data in the passive probe (\Visualization) could contribute to cognitive overload, making it more overwhelming than helpful to interpret AI-generated insights during discussions. They felt the topics presented were too abstract to be meaningful while expanding on them risked consuming time and distracting from the discussion \pid{P1, P2, P5, P10, P13}. 

\begin{smallquote}
\iquote{"Skimming already maxes out the cognitive load. Once you start reading the topic, you are already slightly distracted. I don't think I'm gonna click on this for details during the meeting."} \pid{(P13, in \Visualization)} \end{smallquote}

In contrast, the active probe (\Questioning) posed the opposite challenge. Participants found that nudges sometimes failed to capture critical information or lacked relevance, leading to insufficient support for reflection \pid{(6/15)}.

\begin{smallquote}
\iquote{"I expected the questions to guide us better, but sometimes they were just too generic and not connected to what we needed."} \pid{(P10, in \Questioning)}
\end{smallquote}

\subsubsection{Reflection engagement: Limited engagement vs. excessive disruption}
\label{subsubsec:findings-limitedengagement}

The active probe (\Visualization)'s passive nature meant that participants often overlooked or ignored it, limiting their engagement and reflection. It was described as a background process that did not actively draw attention \pid{(6/15)}. 

\begin{smallquote}
\iquote{"It was more of a non-interactive process happening at the side of the screen. The focus is on driving the discussion rather than paying attention to what's displayed on the screen,"} \pid{(P12, in \Visualization)} \end{smallquote}

On the other hand, some participants perceived the interactive nature of the AI nudges in the active probe (\Questioning) as intrusive, interrupting the natural flow of discussions \pid{(4/15)}, especially when frequent or poorly timed. Others \pid{(5/15)} felt these interruptions created pressure to respond, requiring additional time to refocus and potentially reducing meeting efficiency.

\begin{smallquote}
\iquote{"It actually creates interruptions in the meeting. I felt like I had to answer when the AI questioned something. I'm just not sure whether I want that experience in the meeting."} \pid{(P2, in \Questioning)}
\end{smallquote}

\subsubsection{Reaction to errors: Over-reliance vs. sensitivity to mistakes}
\label{subsubsec:findings-errors}

For the passive probe (\Visualization), participants often accepted the AI-generated outputs as accurate and aligned their understanding without critically evaluating the content, which sometimes led them to overlook discrepancies or nuances. During the interviews, 6 participants initially believed the AI-generated goals accurately reflected their meeting objectives. However, upon further questioning and reassessment, they identified discrepancies in the AI's interpretation.

\begin{smallquote}
\iquote{"I think those goals and topics looks good to me.... (After a while)... Actually, I might want to change the last topic to the first goal rather than the last goal"} \pid{(P4, in \Visualization)}
\end{smallquote}

In contrast, some participants were highly sensitive to AI inaccuracies in the active probe (\Questioning), including both the content's relevance and the timing of the AI's interjections \pid{(6/15)}. When AI nudges were off-target or misaligned with the participants' needs, it caused frustration and led to skepticism about the AI's reliability. This frustration was heightened by the effort required to address or redirect the conversation after an irrelevant nudge. 

\begin{smallquote} \iquote{"Particularly given this was the topic I just covered, I'd be frustrated.  Why is it asking me a wrong question? I think I'd smile when it came up, `Oh, Copilot, you think you know, but no, you're wrong'."} \pid{(P14, in \Questioning)} \end{smallquote}

\subsubsection{Impact on social dynamics and inclusiveness}
\label{subsubsec:findings-social}

Participants raised concerns that both probes might affect social dynamics and participation balance in meetings. In the active probe (\Questioning) condition, some participants, particularly individual contributors, feared being judged by AI and called out for straying off-topic \pid{(6/11)}, thus criticized by the team \pid{(P2, P5, P7, P11)}, which could impact their confidence to share ideas and contribute to the meeting. They worried about the potential of `being flagged as off-track' \pid{(P8)}, `being seen as unprofessional' \pid{(P4)}, or `taking up others' time unnecessarily' \pid{(P2)}.

\begin{smallquote}
    \iquote{"What if I'm in a really big meeting with hundreds of attendees? I would want it to pop up just to me with the feedback."} \pid{(P5, in \Questioning)}
\end{smallquote}

Managers were less concerned about being called out and more focused on effectively initiating the reflection process.  They expressed concerns about the potential for further suppressing the voices of junior team members if they were the ones solely responsible for deciding when to reflect \pid{(3/4)}. Additionally, they noted that leaving the decision to everyone could disrupt meeting flow and management \pid{(2/4)}.

\begin{smallquote}
    \iquote{"My only worry is if I have 10 folks in my meeting and one says vote for group goal discussion, will we go for the reflection or just ignore?" }\pid{(P6, in \Questioning)}
\end{smallquote}

While less prominent, participants also noted that the passive probe (\Visualization)'s transparent nature sometimes led to feelings of exposure, as participants feared that the visual representation of goals or topics and distribution of time could reveal their lack of contribution \pid{(4/11)}.

\begin{smallquote}
\iquote{"Even seeing my ideas not represented or shown less on the screen made me feel like I needed to say more just to keep up."} \pid{(P9, in \Visualization)}
\end{smallquote}

\subsubsection{Adaptability to context}
\label{subsubsec:findings-adapttocontext}

Some participants were concerned about the AI's ability to adapt to diverse meeting contexts, especially those without predefined goals \pid{(7/15)}.
They noted that meetings can range from structured project updates with clear agendas to more fluid brainstorming sessions where goals evolve. In these dynamic settings, they worried that the AI's predefined logic might not capture spontaneous shifts or adapt to tacit goals, particularly in exploratory or open-ended meetings \pid{(5/15)}.

\begin{smallquote}
\iquote{"I wasn’t sure if the AI could handle different meeting formats or if it would just stick to a one-size-fits-all approach."} \pid{(P11)}
\end{smallquote}

\subsubsection{Timing and synchronicity of AI nudges}
\label{subsubsec:findings-timing}

Some participants expressed concerns about the synchronicity of AI-generated content in both probes to nudge meaningful reflection \pid{(4/15)}. In the active probe (\Questioning) condition, questions aimed at discussing unclear meeting goals sometimes appeared after participants had already formulated or discussed those goals due to processing delays.

\begin{smallquote}
\iquote{"By the time the AI showed the question, we just established what we were trying to achieve. It felt a bit late and not as useful."} \pid{(P10, in \Questioning)}
\end{smallquote}

In the passive probe (\Visualization) condition, participants noted that the identified goals were sometimes displayed after the topic had shifted or evolved. This lack of synchronicity limited the AI's ability to promote timely reflection and guide the discussion effectively.

\begin{smallquote}
\iquote{"The AI would show a topic that we had already moved past, so it didn't prompt me to think further. It was more like a summarization than a reflection aid."} \pid{(P6, in \Visualization)}
\end{smallquote}

\section{Finding (RQ3): Designing AI-Assisted meeting reflection support}
\label{sec:findings-design}

Although participants recognized the benefits of AI-assisted reflection, they also shared concerns, highlighting the need for thoughtful design. Their feedback revealed three key design dimensions: \textbf{ Who}, \textbf{When}, and\textbf{ What}—revealing user roles, timing, and content needs. Participants' responses also provided insights into \textbf{How} these design dimensions could be implemented practically, emphasizing adapting content to timing and role, adapting interventions' strength to subjective timing, balancing between democratic input and efficiency, and providing more user control.

\begin{figure}[ht]
    \centering
    \includegraphics[width=\linewidth]{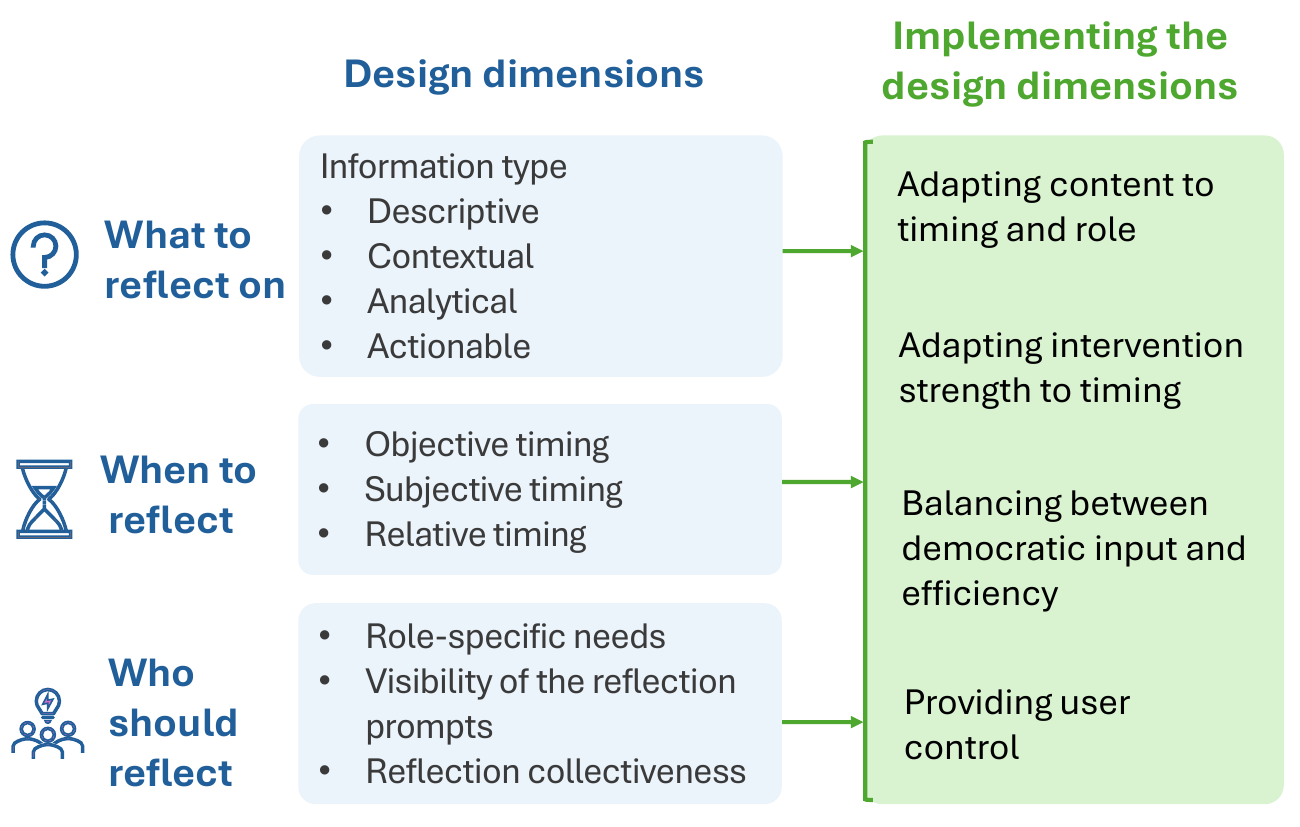}
    \caption{\textit{Design dimensions}---\textbf{what to reflect on, when to reflect, and who should reflect}---and \textit{implementation considerations} for accommodating users' needs for in-meeting reflection. }
    \label{fig:enter-label}
\end{figure}

\subsection{Design dimensions: \textit{What} to reflect on, \textit{when} to reflect, and \textit{who} should reflect}
\label{subsec:findings-design-dimensions}

\subsubsection{What to reflect on: Desired content for reflection}
\label{subsubsec:findings-design-what}

Participants noted that the passive probe (\Visualization) sometimes delivered an overwhelming amount of information, whereas the active probe (\Questioning) sometimes missed important details they wished to consider (§\ref{subsubsec:findings-overload}). Participants expressed the need for information that effectively balances cognitive overload with the necessity for essential, actionable insights during the meeting.
We categorized responses about information needs into four key content areas: descriptive, contextual, analytical, and actionable (\autoref{tab:content-categories}). 

\renewcommand{\arraystretch}{1.25} 
\begin{table*}[h!]
\centering
\begin{tabular}{p{0.1\textwidth}p{0.27\textwidth}p{0.53\textwidth}}
\toprule
\textbf{ Category} & \textbf{Content} & \textbf{Description} \\
\hline
\textbf{Descriptive } & Predefined goals & Goals or objectives identified before or at the start of the meeting. \\
\cline{2-3}
& Pre-defined agendas & The agenda or topics to be discussed during the meeting identified before or at the start of the meeting.\\
\cline{2-3}
& Emergent goals/topics & New goals or topics that arise during the meeting. \\
\hline
\textbf{Contextual} & Async collaboration updates & Information from tasks or discussions completed asynchronously before the meeting. \\
\cline{2-3}
& Prior meeting data & Summary of decisions and action items from previous meetings. \\
\cline{2-3}
& Key metrics/indicators & Relevant data points like KPIs, performance metrics, and deadlines. \\
\cline{2-3}
& Document references & Links or references to documents and materials relevant to the meeting or being discussed during the meetings. \\
\cline{2-3}
& Context-related questions & Questions that promote deeper thinking about the current topic. \\
\cline{2-3}
& Provoking new thoughts & Contextual suggestions that help the team consider new perspectives. \\
\hline
\textbf{Analytical} & Priority and time allocation of topics & Assessment of the importance of topics, suggesting their order and time allocation. \\
\cline{2-3}
& Relation and hierarchy of topics and goals & Mapping how discussion points relate to each other and to overall objectives. \\
\cline{2-3}
& Deviation analysis & Identifying and addressing any deviations from the main goals or objectives. \\
\cline{2-3}
& Personal contribution and team dynamics & Insights on how individual contributions and team interactions are affecting meeting goals. \\
\hline
\textbf{Actionable} & Next topic & Nudge to move to the next item on the agenda. \\
\cline{2-3}
& Decisions, or conflicts to be addressed & Specific decisions that need to be made immediately or conflicts to be addressed. Guide the discussion with necessary action.  \\
\cline{2-3}
& To-do for me & Specific tasks assigned to individuals to complete after the meeting. \\
\cline{2-3}
& Follow-ups for the team & Reminders or questions about actions to be taken after the meeting. \\

\bottomrule
\end{tabular}
\caption{Information categories that are desired by users for AI-assisted in-meeting reflection.}
\label{tab:content-categories}
\end{table*}

\paragraph{Descriptive information.}
Participants emphasized the need to have concise information showing an overview of the meeting without further analysis, including clear and accessible pre-defined goals \pid{(P3, P7, P8, P10, P11, P13, P14)}, agendas \pid{(P1, P2, P4, P6, P9)}, and any emergent goals or unplanned topics that arise during the meeting to address the uncertainty issues \pid{(P3, P11, P12, P14)}. As suggested by them, these elements can help tackle the challenges of unarticulated goals (\S\ref{subsec:findings-agendas}) and poor tracking in external artifacts (\S\ref{subsec:artifacts}) and address uncertainty in goal-setting (\S\ref{subsec:findings-uncertainty}). 

\begin{smallquote}
    \iquote{"My ideal system would as soon as we're in this meeting [pointing to a slide with agenda listed], it would get these topics and stick it on the screen. So you know, here are the meeting goals from the top of your mind."  } \pid{(P14, in \Visualization)}
\end{smallquote}


\paragraph{Contextual information.} Beyond information directly related to a given meeting, participants emphasized the need for broader contextual information to accommodate the varying types and contexts of meetings that influence goal-setting and goal-tracking (\S\ref{subsec:findings-uncertainty}). This included references to information from asynchronous collaboration workflows \pid{(P5, P6, P12)} or records from previous instances of a recurring meeting series \pid{(P1, P3, P4, P7, P8)}.
They also wanted AI to provide context relevant to the ongoing discussion, such as capturing key data points mentioned during meetings \pid{(P1, P6, P10, P12)} or linking relevant mentioned documents \pid{(P7, P9, P14, P15)} to help attendees' sense-making.

\begin{smallquote}
  \iquote{"
  For example, as part of the onboarding, our decision is to reduce this onboarding time to 15 minutes for a customer. I'd expect the metric of 15 minutes to be captured." }  \pid{(P12, in \Visualization)}
\end{smallquote}

Participants suggested that beyond displaying goals and topics, contextual information should also augment discussions \textit{constructively} \pid{(7/15)}, including questions to promote deeper thinking  \pid{(P3, P6, P7, P11)} or suggest new perspectives \pid{(P2, P9,  P12)}. 

\begin{smallquote}
    \iquote{"We're dealing with a lot of security things, and the AI could pop up a question saying that `have you thought about the security context or the security issues with this type of an onboarding'?  
"} \pid{(P6, in \Questioning)}
\end{smallquote}

\paragraph{Analytical information.}
Participants wanted AI to provide analysis beyond data visualization to help them quickly grasp key points \pid{(8/15)}, 
such as prioritizing remaining agenda items for better time management \pid{(P3, P4, P10)}, showing progress towards meeting goals \pid{(P4, P6, P11, P12)}, 
and providing insights into group dynamics and individual contributions \pid{(P2, P6, P11)}. User feedback indicates these features may amplify the advantages of AI-assisted reflection, such as personal sense-making (\S\ref{subsubsec:findings-sensemaking}) and group dynamics (\S\ref{subsubsec:findings-team-dynamics}).

\begin{smallquote}
    \iquote{"I think if it gives a judgment or an estimate of how well the meeting goes based on the assumed topics and goals, and then just asking if should we move on as it goes or should we kind of re-structure"} \pid{(P10, in \Questioning)}
\end{smallquote}

\paragraph{Actionable information.}
Participants also emphasized the need for directly actionable information.  They wanted to know ``what to do next'' rather than just acknowledging the data, which is crucial for tackling the low engagement observed in the passive probe (\S\ref{subsubsec:findings-limitedengagement}) and enhancing the efficacy of AI-assisted reflection actions (\S\ref{subsec:findings-benefits}). Two types of actionable information were mentioned. Information that is \textit{`actionable for now'} refers to nudges on how to proceed for now, e.g., moving to the next topic \pid{(P5, P11, P14)} or discussion points that require a decision to be made \pid{(P1, P2, P9)}. 

\begin{smallquote}
\iquote{"It is not informed enough to only ask the questions. A behavioral prompt would be more useful. AI can pop up to say, `
Are you guys ready to move on to project work?' This is most functional, isn't it?"} \pid{(
P11, in \Questioning)}   
\end{smallquote}

On the other hand, information that is \textit{`actionable for later'} guides post-meeting work and further collaboration, ensuring that both individuals and the team know what tasks to follow up on \pid{(P3, P6, P7, P13)}.

\subsubsection{When to reflect: Objective, subjective, and relative timing}
\label{subsubsec:findings-design-when}

The timing of intervention is another key dimension of designing AI systems for in-meeting reflection.
Participants' responses centered around \textit{objective} timing (e.g., meeting start or end), \textit{subjective} timing (pertaining to `critical moments' or other subjective needs), and \textit{relative} timing (that between the reflection nudge and the triggering content). 

\paragraph{Objective timing}
The start of the meeting was seen as the most important moment for having AI guide people in reflecting on goals for the discussion ahead \pid{(8/15)}. 
As suggested by users, reflection nudges can also be introduced at key moments towards the end of the meeting---but not \textit{too} late to take action---to help ensure the meeting completes successfully \pid{(6/15)}. 

\begin{smallquote}
\iquote{ "I think the question is most useful around the 2/3 point of the meeting when we still have time to adjust but are deep enough into the discussion to know what needs realignment." }  \pid{(P4,  in \Questioning)}
\end{smallquote}

\paragraph{Subjective timing}
All users wanted AI assistance during critical moments in meetings, but the definition of a "critical moment" varied based on individual priorities, concerns, and context, reflecting the concept of subjective timing.

Participants defined `critical moments' as those that lead to negative team outcomes if no intervention occurs 
\pid{(P2, P3, P4, P8, P9)}. This included discussions that veer far off-topic, particularly for disproportionately long periods of time.  

\begin{smallquote}
 \iquote{ 
  "Maybe for the first 3 minutes, it is OK to slightly off track. If this AI just pops out here and everybody has to wait, it feels a little bit too much. But if we're off track for a long time already, it’s a more critical situation, then the AI should just kick in.”} \pid{(P9,  in \Questioning)}
\end{smallquote}

The subjective need for reflection is also shaped by meeting types and team dynamics. For example, decision-making meetings might require more frequent reflection to maintain goal clarity \pid{(P2, P3, P9, P12)}, whereas team or project update meetings may not \pid{(P4, P7, P8, P13, P15)}. 


\paragraph{Relative timing}
Relative timing refers to the interval between the reflection nudge and the triggering context. 
As mentioned in concerns, a reflection nudge that is too late reduces its usefulness and potentially disrupts the meeting flow (\S\ref{subsubsec:findings-timing}).
Participants pointed out that reflection nudges should occur in sync with the triggering discussion to facilitate taking action \pid{(P1, P3, P7, P10)}. 

\begin{smallquote}
    \iquote{
    If there's no delay, I think it's verifying that `AI found out some moments that we want to further talk about`"} \pid{(P1, in \Visualization)}  
\end{smallquote}

\subsubsection{Who should reflect: Roles, and the visibility and collectiveness of reflection}
\label{subsubsec:findings-design-who}

The dimension of `who' should reflect revolves around several considerations: \textit{role-specific needs} (active and passive attendees, and meeting organizers) and the \textit{visibility and collectiveness} of reflection.  

\paragraph{Role-specific needs.}
\label{sec:findings-design-role}

 Participants frequently highlighted how different roles require different content to elicit reflection. We identified three key roles from the data.
\textit{Active} participants are those who frequently speak and contribute content; \textit{passive} participants
mainly listen and provide feedback when necessary; whereas \textit{organizers} manage the meeting’s process, ensuring flow and coordinating participants. 

Active participants need reflection nudges that help them evaluate their contributions \pid{(P2, P4, P7, P9, P12)}. These nudges can help them determine whether they are dominating the discussion \pid{(P2, P12)}, whether they need to adjust their speaking frequency \pid{(P4, P7)}, or whether they are straying from meeting goals \pid{(P9)}.

\begin{smallquote}
    \iquote{'Not the audience, the presenter needs it, `Hey, you're going off track.`'} \pid{(P7,  in \Questioning)}
\end{smallquote}

In contrast, passive participants, often as listeners or recipients of reports in meetings, need contextual information to stay engaged and understand the flow of the meeting \pid{(P1, P2, P6)}. 

\begin{smallquote}
  \iquote{"It's more helpful to him [the person being reported to]  because he may get lost about my experiments." 
  }\pid{ (P1, in \Visualization)}  
\end{smallquote}

Organizers, regardless of their level of participation, have unique needs to monitor the overall meeting flow \pid{(P3, P10, P15)}. They require reflection nudges to track progress toward goals \pid{(P3, P8, P13)}, ensure the meeting stays on schedule \pid{(P5, P12)}, and adjust the agenda as needed \pid{(P3, P10)}.


\paragraph{Visibility of reflection nudges: Private vs. public nudges.}

One key consideration is who should see the reflection nudge. 
Public reflection nudges---visible to the entire team---were appreciated for their ability to keep everyone aligned with the meeting's progress \pid{(P4, P6, P9, P13)}. 
On the other hand, participants mentioned the need to see the reflection nudges privately \pid{(P3, P5, P8, P10, P11, P12)}, allowing them to freely reflect without fear of external judgment and minimize interruption to the meeting, particularly in hierarchical meetings 
(\S\ref{subsubsec:findings-social}).

\begin{smallquote}
    \iquote{"If the message was especially just available to me, I might be like actually `yeah, I agree` whether AI intervenes either wrongly or actually points out the deviation, but since it's private, people won't know that it's me.  "} \pid{ (P3, in \Questioning)}
\end{smallquote}


\paragraph{Collectiveness: Personal vs. team reflection.} 

Furthermore, participants emphasized the distinction between personal and collective reflection. Personal reflection does not necessarily influence team behavior. Passive nudges, such as visualizations or ambient cues seen in the \Visualization \ probe, were generally regarded as encouraging personal reflection unless explicitly directed at the team \pid{(8/15)}.

 \begin{smallquote}
     \iquote{'I could probably be alerted a little bit and then realize that what I am talking about is not something the team is aiming for. I will think about whether I want to quickly wrap it up or I can manually eliminate this alert if it is not important. '} \pid{(P4, in \Questioning)}
 \end{smallquote}
 
Collective reflection involves the entire team assessing the meeting's progress, direction, or decisions. Active nudges (\Questioning) were found to drive team reflection as they require participants to respond (even to dismiss the nudge), nudging the team to pause the ongoing discussion and reflect together \pid{(9/15)}.


\subsection{Implementing the design dimensions}
\label{subsec:findings-design-implementing}


This section expands upon the dimensions above to explore the issues involved in implementing them to accommodate users' needs for in-meeting reflection. Drawing on participants’ feedback, we present four considerations for implementation: adapting content to timing and role, adapting intervention strength to subjective timing, striking a balance between democratic input and efficiency, and providing more user control. These considerations bridge the identified dimensions and provide actionable strategies for implementing effective AI interventions for meeting reflection. 

\subsubsection{Adapting content to timing and role}
\label{subsubsec:findings-design-implementing-contenttotiming}

Participants mentioned striking the right balance between minimizing cognitive load and providing just enough information for effective reflection \pid{(8/15)}. This specificity of content for reflection should be adapted to both the relative \textit{timing} \pid{(P1, P4, P5, P12, P13)} of the reflection nudge and the receiver's \textit{role} \pid{(P1, P3, P9, P11, P13)}. 

Immediately after a topic is discussed, high-level summaries are perceived as sufficient as the information is still fresh and easily processed \pid{P1, P3, P7}. However, when revisiting a topic later in the meeting, more detailed information about key decisions, unresolved questions, or next steps was mentioned as necessary to trigger more effective reflection \pid{(P1, P12, P13)}. 

\begin{smallquote}
    \iquote{"I should be able to understand it for now. But if I want to take a look on what has been discussed later, I'd like to expand on details" } \pid{(P1, in \Visualization)}
\end{smallquote}

The desired content specificity also varied depending on the receiver's role. Users felt active participants of the meeting required less detailed summaries since they were already familiar with the content \pid{(P1, P9)}. In contrast, they thought passive participants, such as observers or stakeholders, may need more detailed information to understand the context of the decisions made \pid{(P1, P3)}. 

\begin{smallquote}
    \iquote{“ I can reflect on the high-level topic because I'm familiar with the details, but I'm afraid that other listeners cannot. 
    If he is going to see it, he needs more information."}
    \pid{(P1, in \Visualization)}
\end{smallquote}

\subsubsection{Adapting intervention strength to timing}
\label{subsubsec:findings-design-implementing-strengthtotiming}

Another emerging theme concerned how AI intervention strength---the intensity and type of support provided---should align with users' varying levels of subjective assistance needs (i.e., subjective timing, as per \S\ref{subsubsec:findings-timing}), throughout a meeting \pid{(12/15)}. 
This offers practical design recommendations to tackle limited engagement in passive reflection and high disruption in active reflection (\S\ref{subsubsec:findings-limitedengagement}). 

Both light ambient and stronger intermittent interventions were helpful when aligned with individuals' needs and timing. Continuous ambient cues, like color changes (\Visualization), were seen as working well for low assistance needs \pid{(P3, P5, P8, P10, P12)}. However, participants noted that when the discussion drifted significantly off track, light, ambient support could be insufficient and potentially ignored if it was too subtle \pid{(P4, P6, P9, P11, P14)}. 

 \begin{smallquote}
    \iquote{“I sometimes overlook those small reminders if I’m deeply involved in the discussion. Sometimes, I need something stronger to really catch my attention.”} \pid{(P11, in \Visualization)}
\end{smallquote}


In contrast, stronger intermittent interventions, such as active reflection questions or alerts about misalignment (\Questioning), were more attention-grabbing but risked disrupting the discussion flow if poorly timed \pid{(P2, P7, P8, P10)}.
Participants suggested that light-weight notifications could serve as a balance between ambient cues and strong interventions 
\pid{(P4, P12)}. 
They also emphasized the need for interventions to adapt to specific moments, such as only providing stronger nudges when discussions deviate significantly off track and maintaining subtle cues during routine \pid{(P7, P8, P10)}



\subsubsection{Striking a balance between democratic input and efficiency}
\label{subsubsec:findings-design-implementing-democraticefficiency}

Participants highlighted the benefits and necessity of team reflection (\S\ref{subsubsec:findings-design-who}),  but it remains unclear how this can be implemented in a way that maintains efficiency while ensuring inclusiveness. Participants suggested strategies for managing collective reflection nudges, including the gating of active nudges by role and mechanisms for initiating reflection.

\paragraph{Gating active nudges by role.}
\label{subsubsec:findings-design-implementing-gatingbyrole}

In cases where the AI is nudging active reflection, meeting organizers or facilitators were often seen as the ones to receive reflection nudges first in order to gate their use and thereby reduce interruption and maintain the meeting's flow \pid{(P3, P5, P6, P10, P11, P14)}. 

\begin{smallquote}
    \iquote{"You would want to provide this information to the meeting organizer so that they could understand and make a decision on whether to spend the time with the team  to reflect"} \pid{(P14, in \Questioning)}
\end{smallquote}
 

\paragraph{Initiating collective reflection.}
\label{subsubsec:findings-design-implementing-collective}

We explored voting as one potential mechanism for initiating collective reflection. Voting was seen to provide a safe space for participants to voice their needs for collective reflection in a democratic and non-intrusive manner \pid{(9/15)}. 
However, voting may not be suitable for all meeting contexts. In large meetings, users are concerned it could lead to chaos or be misused to disrupt the process \pid{(P6, P11)}.
In less active meetings, voting may create awkward situations if participants refrain from voting \pid{(P8, P12)}. Some participants suggested alternative mechanisms, such as an anonymous button to initiate collective reflection. 

\begin{smallquote}
    \iquote{" If there's a button for me to click to suggest, should we talk about what success looks like for this meeting' and other people see it? Maybe that's easier."} \pid{(P12, in \Questioning)}
\end{smallquote}


\subsubsection{Providing user control}
\label{subsubsec:findings-design-implementing-control}

Finally, participants' feedback focused on the interactivity between users and the AI, including considerations about \textit{human control} over the system and \textit{AI explainability}. Participants expressed a desire to have some degree of control over the AI’s functions, such as manually inputting the goals and correcting AI errors \pid{(8/12)}. Moreover, we observed variations in users' preferences, e.g., for information needs and subjective timing. This suggests that the system should provide control over reflection features at the meeting and individual levels. 

\begin{smallquote}
    \iquote{“ I would want to have the ability to decide what the goals and topics are. I would say having that option, but not make it reliant on it.”} \pid{(P10)}
\end{smallquote}  

The ability of the AI to explain its nudges and decisions was found to be crucial for building trust and ensuring that participants are willing to engage with the AI-assisted reflection \pid{(P1, P2, P4, P7, P10)}.

\begin{smallquote}
    \iquote{"I would always like questioning the rationale behind it. 
    So, I guess I don't object to AI asking questions, but I just want to know the reason why. "} \pid{(P4)}
\end{smallquote}

\section{Discussion}
\label{sec:discussion}

Above we have explored how knowledge workers might engage with AI-assisted reflection in meetings (\S\ref{sec:findings-perceptions}). Although our participants appreciated the potential for AI's role in encouraging deliberate actions (\S\ref{subsec:findings-benefits}), they noted issues such as cognitive burden, dependency, disrupted conversation flow, and inclusivity in meetings (\S\ref{subsec:findings-concerns}). We also explored how our participants' feedback informed our identified design dimensions (\S\ref{subsec:findings-design-dimensions}) and raised implementation issues (\S\ref{subsec:findings-design-implementing}). In this discussion, we connect issues of reflection research to the domain of meeting science, by exploring reflection as a deliberate practice in meetings (\S\ref{subsec:discussion-deliberate}). We then connect user practices, probe concerns, and feedback to discuss design trade-offs, providing implications for future AI-assisted meeting reflection design (\S\ref{subsec:discussion-tradeoffs}). Finally, we take a holistic view of meeting intentionality, discussing strategies to enhance intentional behaviors beyond reflection and throughout the meeting lifecycle (\S\ref{subsec:discussion-holistic}).





\subsection{Reflection as a deliberate practice in meetings}
\label{subsec:discussion-deliberate}

Reflection during meetings introduces unique challenges and opportunities compared to reflection in other settings, such as journaling \cite{gao2012design} and learning \cite{wolfbauer2023rebo}, thereby extending the understanding of reflection in practice to a novel usage scenario. Unlike traditional research on reflection in the workplace, which often emphasizes post-event analysis \cite{suchman2020understanding, bentvelzen2022revisiting},
meetings demand real-time \textit{reflection-in-action} that must balance immediate participation with ongoing reflective thought.
Our findings suggest that integrating reflection into meetings may disrupt the flow momentarily, but may ultimately enhance efficiency by aligning participants with their goals (\S\ref{subsec:findings-benefits}). This aligns with "slow design" theories, which advocate for intentional pauses to foster engagement and deeper reflection \cite{grosse2013slow}.

One key finding is that meetings do not always require deep and active reflection. This contrasts with much of the existing literature that promotes deeper reflection as inherently more meaningful \cite{boud2013reflection, bentvelzen2022revisiting}. Given the fast-paced, cognitively demanding nature of meetings \cite{cao2021large, geimer2015meetings}, our findings suggested that lower levels of reflection (e.g., in the passive \Visualization \ probe), though not always motivating immediate action, were also shown to help maintain focus and alignment without overwhelming participants. Conversely, while high-intensity, deeper reflection (e.g., in the active \Questioning \ probe) can drive real-time intentional action, it may impair productivity if not timed appropriately.

Another distinguishing feature of meeting-based reflection is its fluid interplay between individual and collective processes. While much of the literature on reflection in collaborative environments---including both workplace \cite{kocielnik2018designing, fritz2023cultivating} and CSCL contexts \cite{baker1997promoting, lazareva2021reflection, govaerts_visualizing_2010}---focuses on individual reflection or structured group debriefing, our work shows that meetings require reflection to shift dynamically between private and public domains. 

Finally, reflection in meetings requires adaptability to evolving contexts and time-critical demands. While structured scripts or questions can guide reflection in traditional collaborative learning scenarios \cite{dillenbourg2002over, wuReflectiveTutoringFramework2011}, meetings require adaptable reflection practices catering to individual content, time-critical needs, and evolving discussion. 


\subsection{Design trade-offs for AI-assisted reflection during meetings}
\label{subsec:discussion-tradeoffs}

\S\ref{subsec:findings-design-dimensions} outlined participants' thoughts on design dimensions for AI-assisted meeting reflection, and \S\ref{subsec:findings-design-implementing} outlined participants' considerations about implementing those dimensions. Here, we synthesize those design findings with our findings on reflection benefits and concerns (\S\ref{sec:findings-perceptions}). We integrate insights from recent work on GenAI applications and time-critical team support tools and explore how future systems can deliver desirable AI assistance by addressing three key trade-offs: cognitive load vs. timing and roles, engagement vs. interruption, and inclusiveness vs. efficiency.

\subsubsection{Balancing information overload with contextual relevance: Adapting content specificity to timing and role }
\label{subsec:discussion-overload}

Our findings reveal a trade-off between ambient visualizations in the passive probe (\Visualization), which risk overload, and questioning in the active probe (\Questioning), which may lack detail. It suggests AI should deliver the right amount of information at the right time and tailor content to the individuals most in need so as not to overwhelm users while ensuring they receive relevant and useful information to support effective reflection.

Participants noted that the content of reflection nudges should vary with the discussion state:  during active discussions, brief summaries of key topics minimize distraction, while revisiting earlier topics benefits from detailed insights to re-engage participants. 
Our passive probe (\Visualization), allowing users to adjust topic detail, aligns with this need. Systems like MeetMap support multi-level information granularity, from high-level meeting topics to detailed transcripts \cite{chen2025meetmap}. Similarly, tools like WorldScribe leverage LLMs to generate adaptive, context-aware content using multimodal inputs \cite{chang2024worldscribe}. This suggests the potential for AI-assisted meeting reflection to incorporate multimodal inputs, such as transcripts, shared artifacts (e.g., slides), and nonverbal cues, to deliver contextually relevant content.

Our findings also suggest that AI should tailor content to user roles: active contributors benefit from targeted cues to align their input with objectives, while passive participants need more contextual updates to better understand the discussion. This aligns with prior research indicating that post-meeting tools should let users adjust content detail and select specific information when sharing notes \cite{wang2024meeting}.  By mapping the four categories of information (descriptive, contextual, analytical, and actionable) to specific user roles, our findings provide empirical data for future designers and researchers to explore how these categories can be aligned with specific user roles to enhance AI-assisted reflection.
 Future systems could further enhance role-specific content delivery by leveraging user personas and Role-Playing Language Agents (RPLAs) to improve LLM personalization and performance \cite{chen2024persona}.

\subsubsection{Balancing engagement with interruption: Adapting intervention strength to timing}
\label{subsec:discussion-interruption}



\begin{figure*}[htbp]
    \centering
    \includegraphics[width=1\linewidth]{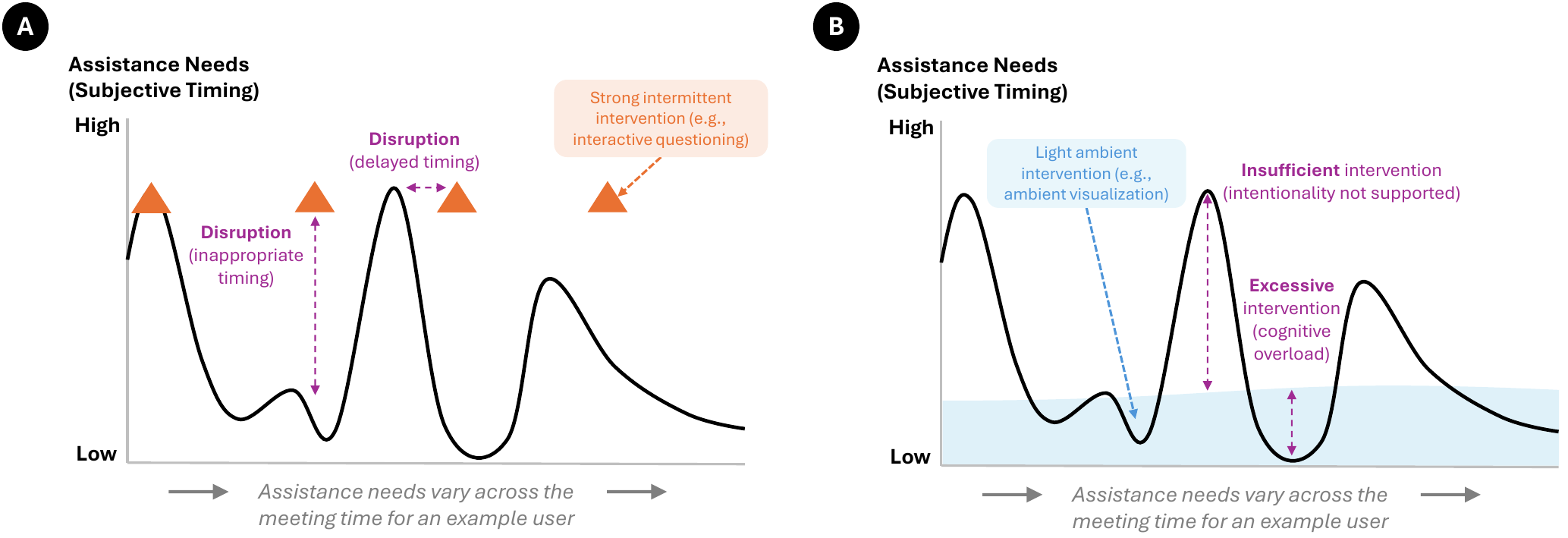}
    \caption{(A) `Strong' \textit{intermittent} interventions are much more direct and can provoke immediate action. However, they risk becoming disruptive if not timed properly. (B) `Light' \textit{ambient} interventions are more subtle. However, when subjective assistance needs are high, they may fail to capture attention. Conversely, when assistance needs are very low, even a light ambient intervention may unnecessarily add to users' cognitive load. }
       \label{fig:comparison-interventions}
\end{figure*}

Our findings emphasize 
that when participants are deeply engaged in discussions, strong interventions can overwhelm and disrupt the flow (\autoref{fig:comparison-interventions}A). Conversely, during moments of confusion or misalignment, subtle interventions may fail to nudge necessary reflection or correction (\autoref{fig:comparison-interventions}B).

\begin{figure} [ht]
    \centering
    \includegraphics[width=1\linewidth]{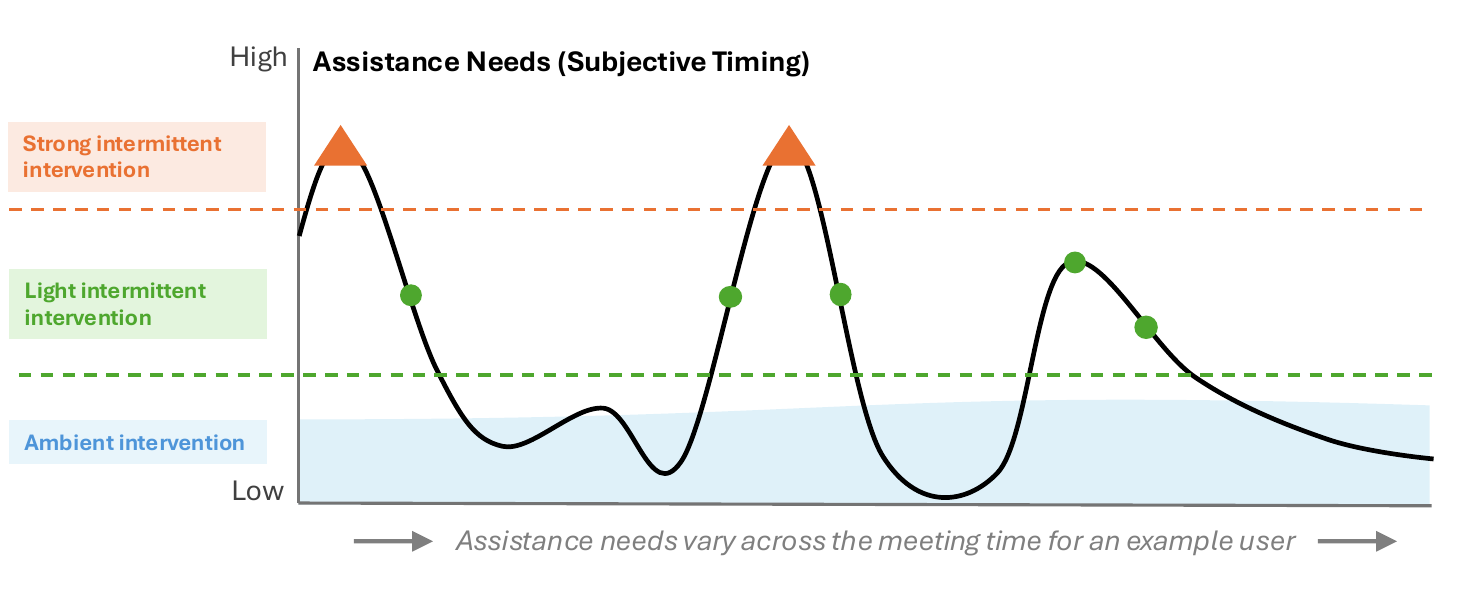}
    \caption{AI should adapt the intervention strength to match users' assistance needs during meetings. }
    \label{fig:desirable}
\end{figure}

We propose \textit{subjective timing} as a lens to determine when participants are in need of support and can accommodate interventions without disruption. AI systems should dynamically adjust intervention intensity between different intervention levels, from ambient through light-weight to strong intervention, escalating or de-escalating the intensity of intervention based on the subjective timing threshold 
(\autoref{fig:desirable}).


A key challenge in designing such systems is determining the appropriate timing (\S\ref{subsubsec:findings-design-when}). Recent studies show the promise of inferring user intentions by analyzing speech \cite{chowdhary2024suggesting}, facial expressions \cite{aseniero2020meetcues}, and meeting interface actions \cite{xia2023crosstalk}. GenAI can further enhance this process by leveraging task-related cues—such as keywords, actions, or user queries—to infer what users aim to achieve, aligning with their needs for actionable information during reflection \cite{rayan2024exploring}.  
In addition to proactive AI assistance, participants in our study also expressed a preference for lightweight interactions to express intentions over intrusive interventions (\S\ref{subsubsec:findings-design-implementing-strengthtotiming}) and sought more control over AI nudges (\S\ref{subsubsec:findings-design-implementing-control}). Aligning proactive AI behavior with user-driven interactions, as recent studies suggest, can enhance the user experience by using feedback loops to deliver timely, context-aware assistance while preserving user autonomy \cite{subramonyam2023bridging}.

Our findings reveal that participants not only have subjective timing needs but also require support at important moments of objective time, such as the beginning or end of meetings (\S\ref{subsubsec:findings-design-when}), highlighting the value of phase-based assistance \cite{park2024coexplorer}. Additionally, participants emphasized the importance of minimizing the relative timing of delays in AI assistance (\S\ref{subsubsec:findings-design-when}). Such real-time need is seen as essential in the high-paced and time-sensitive environment of meetings \cite{mastrianni2023transitioning}. To avoid latency issues, recent studies like MeetMap \cite{chen2025meetmap} advocate for using a temporary palette to present intermediate AI output for real-time sense-making, while the backend continues deeper analysis.

Previous research has suggested enhancing temporal awareness in time-critical teamwork, e.g., in emergency medical teamwork, using concepts including absolute time, relative time, elapsed task time, and time remaining until the next task \cite{kusunoki2015designing}. We argue that meeting temporality should be considered from three perspectives for optimal reflection timing: absolute time (from the meeting's start), relative time (to discussion points), and subjective time (critical moments). These three perspectives should guide the design of AI-assisted reflection tools.

 \subsubsection{Balancing meeting effectiveness with democratic input: Adapting reflection nudge visibility and initiation mechanisms}
 \label{subsec:discussion-democratic}

Our findings suggest that AI-assisted reflection can prompt team actions, such as adjusting agendas. These effects were stronger with the active probe (\Questioning), which required team responses, while the passive probe (\Visualization)  encouraged individual contemplation without directly influencing team behavior. These findings emphasize that the effectiveness of team-wide reflection depends on the visibility of nudges and the initiation of the reflection process.

Participants raised concerns about public nudges, especially in hierarchical meetings, where individual contributors (ICs) feared judgment. Trust levels, team dynamics, and communication styles further influence the value of visible prompts. Previous studies have investigated time-sensitive and hierarchical environments such as operating rooms. In operating rooms, residents are concerned about their attention being visualized, fearing being judged by the attendings \cite{popov2024looking, popov2022towards} and suggest that data display should be on-demand. Similarly, our research highlights the need for AI-assisted reflection tools to balance transparency and privacy by giving users control over sharing their reflections.

When team-wide reflection is needed, determining who initiates it is crucial to maintain meeting flow and inclusiveness.  Our findings suggest the key is to find a balanced mechanism that encourages democratic input while allowing an accountable member, such as the facilitator or organizer, to make the call on whether to initiate collective reflection. 
Previous research highlights assigning an alert owner to clarify responsibility and enhance efficiency in decision-making \cite{mastrianni2022alerts}. One possible solution is to assign one team member to act as the alert owner, which reduces ambiguity about who should react to the alert and facilitates task division \cite{mastrianni2022alerts}. This implies that, although inclusiveness is necessary, in decision-making or urgent meetings, managers or organizers may start reflections to enhance efficiency and meet goals. On the other hand, future research should also explore more inclusive mechanisms for initiating reflections. Our active probe's (\Questioning) voting mechanism was well-received for transparently indicating whether most participants deemed reflection necessary. However, while voting promotes equal participation, it risks overlooking minority voices. Other potential designs, such as anonymous chat \cite{ma2017video} or gathering input to represent participants \cite{leong2024dittos}, could be explored to further ensure that minority voices are heard and encourage shared responsibility. This is especially important for encouraging junior participants to speak up, helping all attendees maintain intentionality and a shared responsibility to make meetings more reflective and efficient \cite{el2019shared}.

\subsection{Beyond reflection during meetings: A holistic view of meeting intentionality}
\label{subsec:discussion-holistic}

Reflection in meetings extends beyond immediate goal alignment, offering broader potential for enhancing collaboration across the entire meeting lifecycle. One notable finding is that participants valued reflection that bridges real-time progress with pre-meeting data and post-meeting outcomes. If these also connect to future meetings, then reflection fosters a cycle between reflection-in-action (where adjustments are made during the meeting) and reflection-on-action  (where insights from the meeting inform post-meeting follow-ups, the next meetings' preparation, and longer-term reviews) \cite{schon2017reflective}. AI can further facilitate this cycle by surfacing pre-meeting goals, guiding in-meeting adjustments, and generating post-meeting summaries. This ongoing process could ensure that intentionality is carried forward not only within individual meetings but across broader project scopes.

\subsection{Limitations and future work}
\label{subsec:discussion-limitations}

Our participants were recruited from one global technology company. While this allowed us to maintain a consistent meeting culture across participants, it limited the diversity of meeting practices. Future research should broaden the scope by including participants from different companies and working areas. 

Additionally, recruitment was limited by confidentiality and ethical considerations, which meant that all attendees in each meeting (including those not participating in the study sessions) had to consent to donating their meeting data for the study. We also had to ensure that no confidential data was included. Thus, our sample of people and work scenarios were limited, and also not gender-balanced. While most findings are shared between individual contributors (ICs) and managers, some role-specific differences emerged, particularly regarding who should initiate reflection. The underrepresentation of managers may have limited the comprehensiveness of capturing all role-based considerations. 
Future work should expand the sample to include diverse participant compositions and explore the dynamics of meetings with varying roles, particularly the contributions of managers and ICs.

Recruiting considerations also influenced the types of meetings we were able to study, although we encouraged participants to reflect on meetings beyond those included in the probes, and also prevented us from deploying the probes during real-time workplace meetings. We mitigated this issue by using real meeting recordings, which increased ecological validity compared to simulated meeting studies, but meant that participants provided retrospective feedback. 
Future research should also investigate deploying AI-assisted reflection tools in live meeting settings to better understand their impact in real-time contexts.

Finally, our probes were only created to explore design dimensions of AI-assisted reflection rather than evaluate GenAI systems. Although we focused on prompting the AI to generate useful outputs, we did not formally evaluate the quality of the AI-generated content. Future research could further test these designs in real meetings, with a focus on evaluating the quality and impact of the AI-generated content in real-time interactions.

\section{Conclusion}
\label{sec:conclusion}

AI-assisted reflection has great potential to enhance intentionality and goal alignment during meetings. However, we found there are trade-offs to be made in how this will be acheived. Reflecting passively seems beneficial for continuous reflection, but might not be able to show its effect in real-time. Active reflection may encourage immediate action, at the risk of disrupting the flow of meetings. Three dimensions are key to supporting reflective goal alignment during meetings: \textit{what} to reflect on, \textit{when} to reflect, and \textit{who} should reflect. Designing for these dimensions will require approaching AI in a new way. Rather than using GenAI as a tool to automate away the mundane procedures of meetings, we believe in developing GenAI as a tool for thought to make meetings more effective. In this case, AI-assisted goal reflection as a deliberate practice may foster intentionality throughout the entire meeting lifecycle. 

\bibliographystyle{ACM-Reference-Format}
\bibliography{Reference}

\appendix

\section{Appendix A: Meeting Data Analysis Code Book}
\label{sec: codes}
\renewcommand{\arraystretch}{1.25} 
\begin{table*}[ht]
\small
\centering
\begin{tabular}{p{2.8cm}p{2cm}p{7cm}p{1cm}p{1cm}}
\hline
\textbf{Category} & \textbf{Code} & \textbf{Definition and Example} & \textbf{ Code Count} & \textbf{Meeting Count} \\ \hline
\multirow{2}{*}{\textbf{Techniques}} 
 & Share agenda in Advance & The practice of distributing the meeting agenda before the meeting in a tangible way, e.g, share in the chat, or put in a shared doc & 5 & 5 \\ 
 & Using Reference & Referring to external materials such as slides or shared screens to keep the discussion focused during the meeting  & 7 & 7 \\ \hline
\multirow{10}{*}{\textbf{Intentional Behaviors}} 
 & Clarify Goals at the Beginning& Ensuring that the goals of the meeting are clearly stated at the beginning.  & 4 & 4 \\ 
 & Clarify Topics at the Beginning & Clarifying the topics or agenda items at the beginning of the meeting. & 8 & 8 \\ 
 & Refer Back to Goals & Revisiting the established goals to check progress or alignment during the meeting.  & 3 & 1 \\ 
 & Refer Back to Topics & Revisiting the specific agenda topics during the meeting to ensure they are being followed.  & 7 & 4 \\ 
 & Prioritization & Determining which topic/tasks should be prioritized during the meeting to maximize efficiency.  & 4 & 2 \\ 
 & Seek Input & Actively encourage participants to contribute their thoughts or feedback.  & 28 & 9 \\ 
 & Delegate Responsibilities & Assigning tasks and setting deadlines to participants during the meeting. & 8 & 4 \\ 
 & Summarize Meeting Outcomes & Summarize the key decisions and action points at the end of the meeting.  & 4 & 2 \\ 
 & Time Management & Monitoring and managing the time spent on each agenda item to ensure timely completion.  & 10 & 5 \\ \hline

\multirow{3}{*}{\textbf{Challenges}} 
 & Discussion Off-Target & When the discussion veers away from the planned agenda or main topic.  & 4 & 3 \\ 
 & Core Issues Not Discussed & Failing to cover important topics due to time constraints or poor agenda management.  & 8 & 4 \\ 
 & Running out of time & Difficulty in managing the time effectively, leading to overtime or unfinished discussions.  & 6 & 6 \\ \hline

\multirow{2}{*}{\textbf{Goal Dynamics}} 
 & Emerging Goals & New goals or objectives that arise unexpectedly during the meeting. & 5 & 3 \\ 
 & Emerging Topics & New discussion topics that were not planned or on the original agenda but arose during the meeting.  & 10 & 7 \\ \hline
\end{tabular}
\caption{Coding Book for Meeting Practices. ("Code Count" refers to the total number of instances a particular behavior was coded, while "Meeting Count" indicates how many distinct meetings featured that behavior at least once.) }
\end{table*}

\end{document}

%% file: _macros.tex
\newtoggle{comments}
\toggletrue{comments}

\iftoggle{comments} {
\newcommand{\lev}[1]{{\color{red}{LT: #1}\normalfont}}
\newcommand{\xinyue}[1]{{\color{magenta}{XC: #1}\normalfont}}
\newcommand{\sean}[1]{{\color{green}{SR: #1}\normalfont}}
}{
  \newcommand {\lev}[1]{}
  \newcommand {\xinyue}[1]{}
  \newcommand {\sean}[1]{}
 }

\newcommand{\pid}[1]{{\fontfamily{cmss}\selectfont{\footnotesize{{\textcolor{black!50}{#1}}}}}}

\newenvironment{smallquote}
  {\begin{list}{}{\setlength{\leftmargin}{1em} 
                  \setlength{\rightmargin}{0em} 
                  \setlength{\topsep}{0em} 
                  \setlength{\partopsep}{0.2em} 
                  \setlength{\parsep}{0em} 
                  \setlength{\itemsep}{0.2em}} 
   \item\normalsize}
  {\end{list}}

\newcommand {\qspace}[1]{\vspace{0.3\baselineskip}}